\documentclass[aps,prb,twocolumn,groupedaddress,amsmath,amssymb,amsfonts]{revtex4}

\usepackage{latexsym,amssymb}
\usepackage{graphicx,color,pstricks}
\usepackage{epsfig,epsf,bm}

\definecolor{gray}{rgb}{0.7,0.7,0.7}

\begin{document}
	
\title{Correlated second-order Dirac semimetals with Coulomb interactions}
	
\author{Yu-Wen Lee}
\email{ywlee@thu.edu.tw}
\affiliation{Department of Applied Physics, Tunghai University, Taichung, Taiwan, R.O.C.}

\author{Yu-Li Lee}
\email{yllee@cc.ncue.edu.tw}
\affiliation{Department of Physics, National Changhua University of Education, Changhua, Taiwan, R.O.C.}
	
\date{\today}
	
\begin{abstract}
 We investigate the effects of long range Coulomb interactions on the low-temperature properties of a 
 second-order Dirac semimetal in terms of the renormalization group. In contrast to the first-order Dirac 
 semimetal, the full rotation symmetry is broken even in the continuum limit, and thus the low-energy 
 physics is controlled by two dimensionless parameters: the dimensionless coupling constant and the ratio 
 of the anisotropy parameters. We show that the former flows to zero and the latter flows to a fixed value 
 at low energies. Thus, one may calculate physical quantities in terms of the renormalized perturbation 
 theory. As an application, we determine the temperature dependence of the specific heat by solving the 
 renormalization-group equations. Following from the breaking of the full rotation symmetry, there exists 
 a crossover temperature scale $T_c$ (and a length scale $L_c$). Physical quantities approach the values 
 for the first-order Dirac semimetal only when the temperature is much smaller than $T_c$. Similarly, the 
 screened Coulomb potential will become anisotropic when the distance is smaller than $L_c$, while the 
 unscreened form is recovered at its tail.
\end{abstract}
	
\maketitle
	
\section{Introduction}

In the past $15$ years, topological phases of matters have become an active research topic in condensed 
matter physics. Among these phases, there are gapped systems, such as topological insulators (TIs) and
topological superconductors (TSCs)\cite{Hasan,Zhang,Ryu}, and gapless systems such as Weyl semimetals 
(WSMs)\cite{XLQi,Vishwanath} and various variants, e.g., anisotropic Weyl fermions\cite{Yang}, double 
WSMs\cite{GXu,CFang}, and tilted WSMs\cite{Soluyanov}. For the TIs, the bulk topology are usually 
indicated by some topological indices, and it is characterized by the monopole charges in the momentum 
space for the WSMs. In these systems, the bulk topology is protected by some (global) symmetries\cite{Ryu}, 
and thus are dubbed as the symmetry protected topological (SPT) phases. For these samples with boundaries, 
there are gapless states localized at the edges, known as the bulk-boundary correspondence. These gapless 
edge states is stable against symmetry-preserving perturbations. 

Later, the study of topological phases was extended to higher order and the previous mentioned TIs now 
belong to the class of the first order\cite{Bernevig1,Wan,Bernevig2}. One way to distinguish the TIs 
belonging to different orders is through their edge states. For example, a Chern insulator, which is a 
type of first-order TIs characterized by the Chern number or quantized Hall conductance, has gapless 
chiral states at its edge. On the other hand, the edge states are still gapped for a second-order TI in 
two dimensions. However, if the boundaries of the sample are compatible with the protecting symmetry, 
then gapless states will exist at the corners. The robustness of these gapless corner states is 
guaranteed by the second-order bulk topology. 

Recently, the notion of higher-order topology has been extended to the Dirac semimetals 
(DSMs)\cite{Hughes1,Roy1,Roy2} as well as WSMs\cite{Wang,Hughes2}. A Dirac point in three dimensions can 
be understood as two degenerate Weyl nodes with opposite monopole charges, where crystalline symmetries 
forbid the two Weyl nodes from hybridizing and opening a gap at each Dirac point\cite{Kane}. Given this 
picture of the bulk, a three-dimensional ($3$D) DSM has two copies of arc-like surface states which 
resemble Fermi-arcs in WSMs, i.e., the double Fermi arcs. However, unlike the surface states in the WSM, 
the double Fermi arcs on the surface of a DSM are not topologically protected. They can be continuously 
deformed into a closed Fermi contour without any symmetry breaking or bulk phase transition\cite{YMLu}. 
Therefore, they are not topological consequences of the bulk Dirac points themselves. In contrast, the
higher-order DSMs (HODSMs) can exhibit robust and nontrivial topology with spectroscopic consequences.
It has been proposed that the HODSM can be realized in the room- ($\alpha$) and intermediate-temperature 
($\alpha^{\prime\prime}$) phases of Cd$_3$As$_2$, KMgBi, and rutile structure ($\beta^{\prime}$-) 
PiO$_2$\cite{Bernevig3,ZMLiao}.

The minimal model of HODSMs contains a pair of Dirac points located at the same energy. For simplicity, 
we take the two Dirac points at the crystal momenta $\pm\bm{K}$ where $\bm{K}=(0,0,k_0)$ with 
$0<k_0<\pi$ (by setting the lattice constant to be unit). Each plane with given $k_z\neq\pm k_0$ in the 
first BZ describes the band structure of a two-dimensional ($2$D) insulator. The two planes at 
$k_z=\pm k_0$ then separate the first BZ into two regions, each consisting of $2$D insulators by treating 
$k_z$ as a free parameter. In a second-order DSM, one of the regions consists of $2$D second-order TIs, 
while the other consists of trivial insulators. For a cubic sample with open boundary conditions in both 
the $x$ and $y$ directions and periodic boundary conditions in the $z$ direction, each $2$D second-order 
TI will host gapless states at its corners. (That is, the energy of the corner states coincides with that 
of the Dirac points.) Hence, these gapless corner states form Fermi arcs on the $1$D hinges terminated at 
the projections of the two Dirac points. It is clear that these Fermi arcs are direct, topological 
consequences of the bulk Dirac points, in contrast with the first-order DSM.

In the present work, we study the effects of the long range Coulomb interaction on a second-order DSM. 
The Coulomb interaction is ubiquitous in condensed matter systems. Especially, its long range nature 
remains intact in semimetals, in contrast with the Fermi liquid. For the $3$D first-order DSM at weak 
coupling, one-loop renormalization group (RG) analysis indicates that the Fermi velocity grows to infinity 
logarithmically at low energies, leading to the marginally irrelevant dimensionless coupling constant\cite{Goswami,Hosur,Sarma}. 
For the second-order DSM, the full rotation symmetry is lost even in the continuum limit, and thus the 
low-energy physics is described by two parameters: the dimensionless coupling constant and the ratio of
the anisotropy parameters. We employ the method of RG and show that the coupling constant flows to zero 
at low energies, similar to the first-order DSM, where as the ratio of the ansiotropy parameters flows to
a fixed value at low energies. To illustrate the effects of the Coulomb interaction, we calculate the 
temperature dependence of the specific heat in terms of the RG equations. We find that in comparison 
with the specific heat of a non-interacting second-order DSM, the Coulomb interaction suppresses its 
value at all temperatures, except at zero temperature and the temperature close to the band width.

The breaking of the full rotation symmetry introduces a crossover temperature scale $T_c$ (and thus a 
characteristic length scale $L_c$). When the temperature is smaller than $T_c$, the values of physical 
quantities are expected to approach the ones for a first-order DSM. However, we find that this 
expectation is realized only at extremely low temperatures or long distances. We illustrate this point by 
the temperature dependence of the specific heat and the momentum dependence of the screened Coulomb 
potential within the RPA approximation.

The rest of the work is organized as follows. We introduce the model of the second-order DSM in Sec. II.
The results of the RPA approximation and the one-loop RG are presented in Sec. III and IV, respectively.
The last section is our conclusions. The details of the calculations are shown in two appendices.

\section{The model}
\subsection{The Hamiltonian}

We start with a lattice model describing the second-order DSM. Its Bloch Hamiltonian is written as 
$H(\bm{k})=H_0(\bm{k})+H_1(\bm{k})$ where\cite{Roy1,Roy2}
\begin{eqnarray}
 H_0(\bm{k}) \! \! &=& \! \! t \! \sum_{j=1,2}S_j\Gamma_j-t_3 \! \left[C_3+\! \sum_{j=1,2}(1-C_j)\right] 
 \! \Gamma_3 \ , \nonumber \\
 H_1(\bm{k}) \! \! &=& \! \! t_1[(C_1-C_2)\Gamma_4+S_1S_2\Gamma_5] \ . \label{hodsmh1}
\end{eqnarray}
Here $C_j=\cos{(k_ja_0)}$, $S_j=\sin{(k_ja_0)}$, $a_0$ is the lattice constant, and $\bm{k}=(k_1,k_2,k_3)$.
The five rank-$4$ $\Gamma$ matrices take the forms $\Gamma_1=\tau_1\otimes\sigma_3$, 
$\Gamma_2=\tau_2\otimes\sigma_0$, $\Gamma_3=\tau_3\otimes\sigma_0$, $\Gamma_4=\tau_1\otimes\sigma_1$, and 
$\Gamma_5=\tau_1\otimes\sigma_2$. One may verify that they are all Hermitian and satisfy the Clifford 
algebra
\begin{equation}
 \{\Gamma_i,\Gamma_j\}=2\delta_{ij} \ , \label{gamma11}
\end{equation}
for $i,j=1,\cdots,5$. Here $\sigma_{\mu}$ and $\tau_{\mu}$ with $\mu=0,1,2,3$ are the standard Pauli 
matrices acting on the spin and orbital spaces, respectively. 

In this model, the two orbitals are the eigenstates of the space inversion $\hat{P}=\tau_3\otimes\sigma_0$. 
On the other hand, the time reversal is implemented by the usual one for spin-$1/2$ fermions, i.e., 
$\hat{T}=\tau_0\otimes(i\sigma_2)\mathcal{K}$ where $\mathcal{K}$ takes the complex conjugation. It is 
straightforward to verify that $H_0(\bm{k})$ preserves both the space inversion (P) and time-reversal (T) 
symmetries, while $H_1(\bm{k})$ breaks both symmetries. However, $H(\bm{k})$ has a PT symmetry, i.e.,
\begin{equation}
 \hat{\Theta}H(\bm{k})\hat{\Theta}^{-1}=H(\bm{k}) \ , \label{hodsmsym16}
\end{equation}
where $\hat{\Theta}\equiv\hat{P}\hat{T}$. Note that $\hat{\Theta}^2=-1$. Therefore, this symmetry assures 
that each energy level labeled by $\bm{k}$ is doubly degenerate. 

The band structure is given by $E_{\pm}(\bm{k})=\pm E(\bm{k})$ where
\begin{equation}
 E(\bm{k})=\sqrt{\! \sum_{j=1}^5[g_j(\bm{k})]^2} \ , \label{hodsmh11}
\end{equation}
where $g_1(\bm{k})=tS_1$, $g_2(\bm{k})=tS_2$, $g_3(\bm{k})=-t_3(C_3+2-C_1-C_2)$, $g_4(\bm{k})=t_1(C_1-C_2)$, 
and $g_5(\bm{k})=t_1S_1S_2$. Each band is two-fold degenerate, as required by the PT symmetry. The upper 
and lower bands touch each other at the points where $E(\bm{k})=0$ or $g_j(\bm{k})=0$ for $j=1,\cdots,5$, 
which are the Dirac points. In this case, the Dirac points are located at $\pm\bm{K}$ where 
$\bm{K}=(0,0,\pi/2)$\cite{foot1}. Notice that the locations of the Dirac points are not affected by 
$H_1(\bm{k})$.

When the Fermi level coincides with the energy at the Dirac points, i.e., $\mu=0$, then the low-energy 
effective Hamiltonian can be obtained by expanding around the two Dirac points, yielding
\begin{equation}
 H_{eff}(\bm{p})=\tilde{h}_+(\bm{p})+\tilde{h}_-(\bm{p}) \ , \label{hodsmh12}
\end{equation}
to the leading terms in $\bm{p}$, where $\bm{k}=\pm\bm{K}+\bm{p}$ and
\begin{equation}
 \tilde{h}_{\pm}(\bm{p})=v_{\perp} \! \sum_{j=1,2}p_j\Gamma_j\pm v_3p_3\Gamma_3+b \! \sum_{j=4,5}
 d_j(\bm{p})\Gamma_j \ , \label{hodsmh13}
\end{equation}
with $v_{\perp}=ta_0$, $v_3=t_3a_0$, $b=t_1a_0^2/2$, $d_4(\bm{p})=p_2^2-p_1^2$, and $d_5(\bm{p})=2p_1p_2$.
Note that $H_{eff}(\bm{p})$ and $H(\bm{k})$ share the same symmetries. Moreover, $b=0$ and $b\neq 0$ 
correspond to the first-order and second-order DSMs, respectively\cite{foot2}.

Now we add the Coulomb interaction to the system. The low-energy effective Hamiltonian around the two 
Dirac nodes $\pm\bm{K}$ are given by
\begin{equation}
 H=\! \sum_{\xi=\pm} \! \int \! \! d^3x\psi^{\dagger}_{\xi}h_{\xi}\psi_{\xi}+H_c \ , \label{hodsmch1}
\end{equation} 
where $h_{\xi}$ is the inverse Fourier transform of $\tilde{h}_{\xi}(\bm{p})$ and
\begin{equation}
 H_c=\frac{1}{2} \! \int \! \! d^3xd^3y\rho_0(\bm{x})V_c(|\bm{x}-\bm{y}|)\rho_0(\bm{y}) \ .
 \label{hodsmch12}
\end{equation}
Here $\xi$ is the node index, $\psi_{\xi}$ and $\psi^{\dagger}_{\xi}$ obey the canonical anticommutation 
relations, $\rho_0(\bm{r})=\sum_{\xi}\psi^{\dagger}_{\xi}\psi_{\xi}(\bm{r})$ is the uniform component of 
the electron density, and $V_c(\bm{r})=e^2/(4\pi\epsilon r)$ with the dielectric constant $\epsilon$.

The action in the imaginary-time formulation is then of the form
\begin{eqnarray}
 S &=& \! \sum_{\xi=\pm} \! \int_X \! \! \psi^{\dagger}_{\xi}(\partial_{\tau}+h_{\xi})
 \psi_{\xi}+i \! \int_X \! \! \phi\rho_0 \nonumber \\
 & & +\frac{1}{2} \! \int_X \! \left[\frac{1}{g_{\perp}^2}|\bm{\nabla}_{\perp}\phi|^2+\frac{1}{g_3^2}
 |\partial_3\phi|^2\right] , \label{hodsms1}
\end{eqnarray}
where $X=(\tau,\bm{r})$, $\int_X=\! \int \! \! d\tau d^3x$, $\bm{\nabla}_{\perp}=(\partial_1,\partial_2)$, 
and $g_{\perp}^2=g_3^2=g^2=e^2/\epsilon$. We have introduced a real bosonic field $\phi$ to describe the 
Coulomb interaction. As usual, the full rotation symmetry is recovered in the continuum limit. In the 
present case, it is not the case when $b\neq 0$ even if we set $v_{\perp}=v_3$. As a result, the 
longitudinal and transverse components of $|\bm{\nabla}\phi|^2$ will acquire different renormalization. 
Hence, we introduce different coupling constants $g_{\perp}^2$ and $g_3^2$ though they have the same bare 
value.

\subsection{The specific heat of the non-interacting DSM}

In the present case, one consequence of the breaking of the full rotation symmetry is the introduction of
a crossover temperature $T_c$. To show this, we first compute the specific heat for the non-interacting 
second-order DSM.

For the non-interacting second-order DSM, the specific heat is given by
\begin{equation}
 c_v(T)=\frac{T^3}{2v_{\perp}^2v_3\pi^2} \! \int^{+\infty}_0 \! dx\frac{g(Tx/T_c)x^4e^{-x}}{(1+e^{-x})^2} 
 \ , \label{hodsmcv1}
\end{equation}
where $T_c=v_{\perp}^2/|b|$ and
\begin{eqnarray*}
 g(x) \! \! &=& \! \! \frac{1}{\sqrt{|x|}} \! \int^1_{w(x)} \! \! \frac{dt}
 {\sqrt{1-t}\sqrt{|x|t-\sqrt{1-t}}} \ , \\
 w(x) \! \! &=& \! \! \frac{2}{\sqrt{1+4x^2}+1} \ .
\end{eqnarray*}
The function $g(x)$ is in a certain sense a measure of the density of states (DOS) for the second-order 
DSM since the latter can be written as
\begin{eqnarray*}
 \rho(\epsilon)=\frac{g(\epsilon/T_c)\epsilon^2}{4v_{\perp}^2v_3\pi^2} \ .
\end{eqnarray*}
When $T\ll T_c$, $g(x)\approx 4$ and we find that 
$\rho(\epsilon)\approx\rho_1(\epsilon)=\epsilon^2/(v_{\perp}^2v_3\pi^2)$, which is the DOS for the 
first-order DSM, and
\begin{equation}
 c_v(T)\approx c_1(T)\equiv\frac{7\pi^2T^3}{15v_{\perp}^2v_3} \ , \label{hodsmcv10}
\end{equation}
which is the specific heat of the first-order DSM.

\begin{figure}
\begin{center}
 \includegraphics[width=0.99\columnwidth]{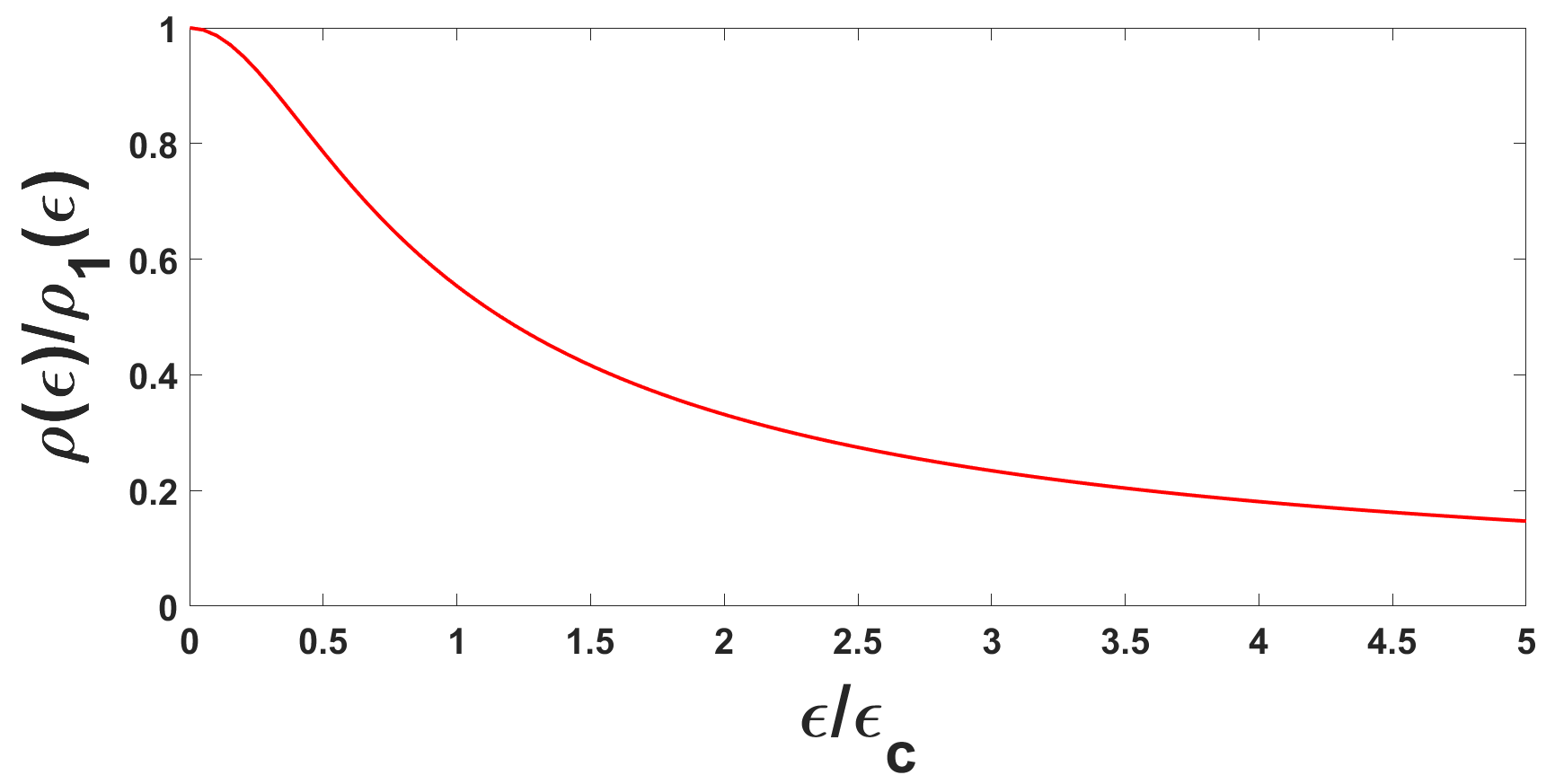}
 \caption{The DOS $\rho(\epsilon)$ as a function of $\epsilon$. We measure $\rho(\epsilon)$ in units of 
 $\rho_1(\epsilon)$, the DOS for the non-interacting first-order DSM, and $\epsilon$ in units of 
 $\epsilon_c=T_c$.}
 \label{hodsmcvf12} 
\end{center}
\end{figure}

Figure \ref{hodsmcvf12} shows the DOS for the non-interacting second-order DSM. We see that it approaches 
the one for the non-interacting first-order DSM only at extremely low energies. $b\neq 0$ suppresses the
value of $\rho(\epsilon)$ compared with the one for the non-interacting first-order DSM. Following from
this fact, in the most temperature range, the thermodynamic properties of a second-order DSM will deviate 
significantly from those of a first-order DSM.

\section{The RPA approximation}

There exists a characteristic length scale $L_c$ associated with the crossover temperature $T_c$ 
introduced by $b$. The momentum or position dependence of a physical quantity will depends on $L_c$ in 
a nontrivial way. To illustrate this point, we consider the screened Coulomb potential. In the momentum
space, its Fourier transform $\mathcal{V}_s(\bm{q})$ satisfies the Dyson equation
\begin{equation}
 \frac{1}{\mathcal{V}_s(\bm{q})}=\frac{1}{\mathcal{V}_0(\bm{q})}+\Pi(0,\bm{q}) \ , \label{scoul1}
\end{equation}
where $\mathcal{V}_0(\bm{q})=g^2/\bm{q}^2$ is the Fourier transform of the bare Coulomb potential 
$V_c(\bm{r})$ and $\Pi(0,\bm{q})$ is the vacuum polarization at zero frequency.

Within the RPA approximation, the vacuum polarization is given by
\begin{eqnarray*}
 & & \! \! \! \Pi(Q) \\
 & & \! \! \! =-\frac{1}{v_{\perp}^2v_3} \! \sum_{\xi} \! \int^{\prime} \! \! \frac{d^3\tilde{p}}
 {(2\pi)^3} \! \int^{+\infty}_{-\infty} \! \frac{dp_0}{2\pi}\mbox{tr} \! \left[G_{\xi 0}(P)G_{\xi 0}(P+Q)
 \right] ,
\end{eqnarray*}
where $Q=(iq_0,\bm{q})$, $\tilde{p}_{1/2}=v_{\perp}p_{1/2}$, $\tilde{p}_3=v_3p_3$, 
$\tilde{b}=b/v_{\perp}^2$, $\tilde{\bm{p}}_{\perp}=(\tilde{p}_1,\tilde{p}_2)$, and
\begin{eqnarray*}
 & & \! \! \! G_{\xi 0}(P) \\
 & & \! \! \! =
 -\frac{ip_0+\! \sum_{j=1,2}\tilde{p}_j\Gamma_j+\xi\tilde{p}_3\Gamma_3+\tilde{b}\sum_{j=4,5}d_j(\tilde{\bm{p}}_{\perp})\Gamma_j}
 {p_0^2+\tilde{\bm{p}}_{\perp}^2+\tilde{p}_3^2+\tilde{b}^2\tilde{\bm{p}}_{\perp}^4} \ , 
\end{eqnarray*}
is the free propagator of the fermion fields. Due to the complicated form of $\Pi(0,\bm{q})$, we will 
evaluate it in two situations: (i) $\bm{q}_{\perp}=0$ and $q_3\neq 0$ and (ii) $\bm{q}_{\perp}\neq 0$ 
and $q_3=0$. We just present the results and discuss their consequences. The details of calculations are 
left to appendix \ref{a1}.

\begin{figure}
\begin{center}
 \includegraphics[width=0.99\columnwidth]{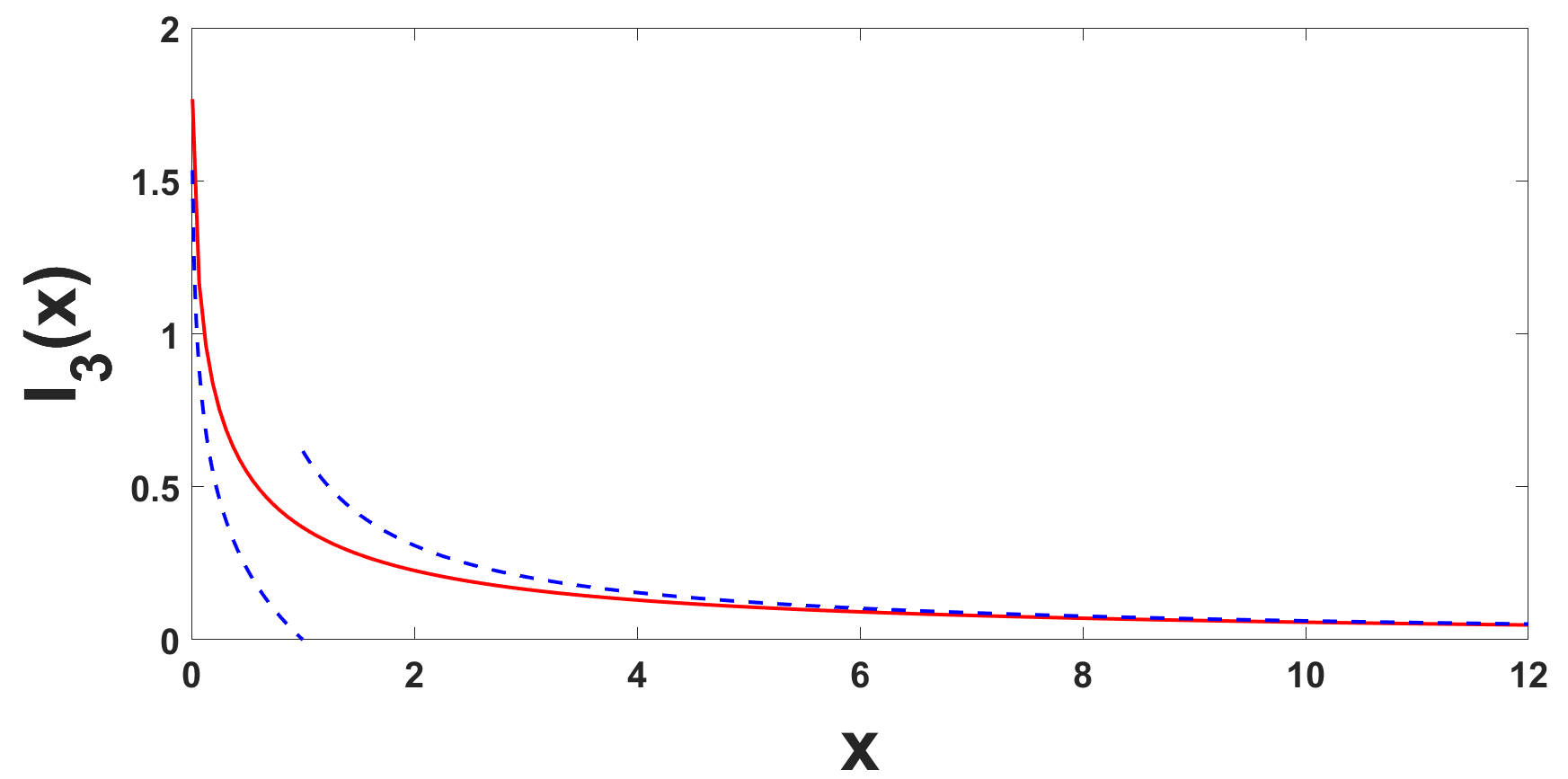}
 \caption{The plot of the function $I_3(x)$ (solid line). The dashed line is the plot of the function 
 $-\ln{|x|}/3$ and $\pi^2/(16|x|)$ for $|x|<1$ and $|x|>1$, respectively.}
 \label{hodsmvcf1} 
\end{center}
\end{figure}

For case (i), we find that
\begin{equation}
 \Pi(0,0,q_3)=\frac{\tilde{q}_3^2}{v_{\perp}^2v_3\pi^2}I_3(\tilde{b}\tilde{q}_3) \ , \label{hodsmvc22}
\end{equation}
where  $\tilde{q}_3=v_3q_3$,
\begin{eqnarray*}
 I_3(x)=\! \int^{+\infty}_0 \! ds \! \left(1-\frac{[\tilde{a}(s,x)]^2-1}{2\tilde{a}(s,x)}
 \ln{\! \left|\frac{\tilde{a}(s,x)+1}{\tilde{a}(s,x)-1}\right|}\right) , 
\end{eqnarray*}
and
\begin{eqnarray*}
 \tilde{a}(s,x)=\sqrt{1+4s+4x^2s^2} \ .
\end{eqnarray*}
We plot the function $I_3(x)$ in Fig. \ref{hodsmvcf1}. One may obtain the asymptotic behaviors of it in 
the limits $|x|\ll 1$ and $|x|\gg 1$:
\begin{eqnarray*}
 I_3(x)=\! \left\{\begin{array}{cc}
 -\frac{1}{3}\ln{|x|} & |x|\ll 1 \\
 & \\
 \frac{\pi^2}{16|x|} & |x|\gg 1
 \end{array}\right. .
\end{eqnarray*}
Thus, in the limit $|\tilde{b}\tilde{q}_3|=L_c|q_3|\ll 1$, we find that
\begin{equation}
 \Pi(0,0,q_3)\approx-\frac{\tilde{q}_3^2}{3v_{\perp}^2v_3\pi^2}\ln{|\tilde{b}\tilde{q}_3|}
 =-\frac{v_3q_3^2}{3v_{\perp}^2\pi^2}\ln{(L_c|q_3|)} \ , \label{hodsmvc23}
\end{equation}
and
\begin{equation}
 \Pi(0,0,q_3)\approx\frac{|\tilde{q}_3|}{16|\tilde{b}|v_{\perp}^2v_3}=\frac{v_3|q_3|}{16L_cv_{\perp}^2} 
 \ , \label{hodsmvc24}
\end{equation}
for $|\tilde{b}\tilde{q}_3|=L_c|q_3|\gg 1$, where $L_c=v_3|b|/v_{\perp}^2=v_3/T_c$.

For case (ii), we find that
\begin{eqnarray*}
 \Pi(0,\bm{q}_{\perp},0)=\frac{\tilde{q}_{\perp}^2}{3v_{\perp}^2v_3\pi^2}
 \ln{\! \left(\frac{\Lambda}{\mu}\right)} \! +\cdots \ , 
\end{eqnarray*}
when $|\tilde{b}|\tilde{q}_{\perp}\ll 1$, and
\begin{eqnarray*}
 \Pi(0,\bm{q}_{\perp},0)\approx\frac{4\tilde{q}_{\perp}^2}{3v_{\perp}^2v_3\pi^2}
 \ln{\! \left(\frac{\Lambda}{\mu}\right)} \! +\cdots \ , 
\end{eqnarray*}
when $|\tilde{b}|\tilde{q}_{\perp}\gg 1$, where $\tilde{q}_{1/2}=v_{\perp}q_{1/2}$, 
$\tilde{\bm{q}}_{\perp}=(\tilde{q}_1,\tilde{q}_2)$, $\Lambda$ is an UV cutoff in $|\tilde{\bm{p}}_{\perp}|$, 
$\mu$ is an IR energy scale, and $\cdots$ denotes the finite part.  In the above calculation for the 
momentum integrals, we follow the strategy employed in Ref. \onlinecite{Yang} and integrate 
$\tilde{\bm{p}}_{\perp}$ and $\tilde{p}_3$ over the ranges $0<|\tilde{\bm{p}}_{\perp}|<\Lambda$ and 
$-\infty<\tilde{p}_3<+\infty$, respectively.

We see that the leading behaviors in $\Pi(0,\bm{q}_{\perp},0)$ are $\Pi(0,\bm{q}_{\perp},0)\sim q_{\perp}^2$ 
in both regimes $q_{\perp}\ll 1/L_c$ and $q_{\perp}\gg 1/L_c$. The logarithmic divergence in front of the 
$q_{\perp}^2$ term can be absorbed into $g^2$, leading to the (infinite) renormalization of the coupling 
constant $g^2$ within the RPA (or large $N$) approximation. (If we replace $\psi_{\xi}$ by $\psi_{\xi\beta}$ 
with $\beta=1,2,\cdots,N$, then the RPA result is exact in the large $N$ limit.) The regimes 
$q_{\perp}\ll 1/L_c$ and $q_{\perp}\gg 1/L_c$ correspond to the first-order and second-order DSMs, 
respectively. Therefore, the different coefficients in front of $q_{\perp}^2\ln{(\Lambda/\mu)}$ imply 
different renormalization of $g^2$ in both regimes. 

In view of Eq. (\ref{scoul1}), the coupling constant $g_{\perp}^2(\mu)$ at the scale $\mu$ can be defined 
as
\begin{eqnarray*}
 \frac{1}{g_{\perp}^2(\mu)}=\frac{1}{g_{\perp}^2(\Lambda)}+\frac{c}{3v_3\pi^2}
 \ln{\! \left(\frac{\Lambda}{\mu}\right)} ,
\end{eqnarray*}
where $c=1,4$ for $q_{\perp}\ll 1/L_c$ and $q_{\perp}\gg 1/L_c$, respectively. Thus, we find that
\begin{equation}
 \Lambda\frac{\partial\alpha}{\partial\Lambda}=\frac{4c}{3}\alpha^2 \ , \label{hodsmrg25}
\end{equation}
where $\alpha=g_{\perp}^2/(4\pi^2v_3)$ is the dimensionless coupling constant. Equation (\ref{hodsmrg25}) 
indicates that $\alpha$ flows to zero at low energies. Accordingly, the DSM is stable against the 
presence of the Coulomb interaction in the large N limit.

The above analysis suggests the following behavior for $\mathcal{V}_s(\bm{q})$. It acquires only a 
logarithmic correction at small momenta, as in the case of the first-order DSM. On the other hand, for 
$q\gg 1/L_c$, 
\begin{eqnarray*}
 \mathcal{V}_s(\bm{q})\approx\frac{g^2}{\bm{q}^2+w|q_3|} \ ,
\end{eqnarray*}
where $w>0$ is a nonuniversal constant. Within the RPA approximation, we get $w=v_3g^2/(16L_cv_{\perp}^2)$.
When $L_c$ is very large, the screened Coulomb potential in the real space is approximately given by
\begin{eqnarray*}
 V_s(\bm{r})\approx \! \int \! \! \frac{d^3p}{(2\pi)^3}e^{i\bm{p}\cdot\bm{r}}\frac{g^2}{\bm{p}^2+w|p_3|} \ ,
\end{eqnarray*}
except its tail. By performing the momentum integral, we find that
\begin{equation}
 V_s(\bm{r}_{\perp},0)=\! \left\{\begin{array}{cc}
 \frac{g^2}{8\pi^2r_{\perp}} & r_{\perp}\ll 1/w \\
 & \\
 \frac{g^2}{2\pi^3wr_{\perp}^2} & r_{\perp}\gg 1/w 
 \end{array}\right. , \label{hodsmvc11}
\end{equation}
where $\bm{r}_{\perp}=(x,y)$ and
\begin{equation}
 V_s(0,z)\approx\frac{g^2}{4\pi c|z|} \ , \label{hodsmvc12}
\end{equation}
for both $|z|\ll 1/w$ and $|z|\gg 1/w$, and $c=1,2$ when $|z|\ll 1/w$ and $|z|\gg 1/w$, respectively. 
(The details of the calculations are left to appendix \ref{a1}.) Therefore, if we place a charged 
impurity at the origin, the induced charge density $\rho_{ind}(\bm{r})$ will be anisotropic in the 
intermediate distance, while its tail ($r\gg L_c$) is approximately isotropic as in the case of the 
first-order DSM.

\section{The renormalization group}
\subsection{One-loop RG equations}

Now we study the effects of the Coulomb interaction in terms of the RG. To perform the RG transformation, 
we decompose each field into the slow and fast modes: $\psi_{\xi}=\psi_{\xi<}+\psi_{\xi >}$ and 
$\phi=\phi_<+\phi_>$ where the subscripts $<$ and $>$ denote the slow and fast modes, respectively. Due
to the lack of the full rotation symmetry, we adopt the regularization scheme employed in the anisotropic
Weyl fermions\cite{Yang} and double WSMs\cite{Lai,Jian}. That is, the fast modes consist of 
the Fourier components between the cylindrical shell $\Lambda/s<|\tilde{\bm{p}}_{\perp}|<\Lambda$ and 
$|\tilde{p}_3|<+\infty$ where $s=e^l>1$.

Now we integrate out the fast modes to the one-loop order, and then perform the scaling transformation
\begin{eqnarray*}
 & & \! \! x(y)\rightarrow e^lx(y) \ , ~z\rightarrow e^{z_3l}z \ , ~\tau\rightarrow e^{zl}\tau \ , \\
 & & \! \! \psi_{\xi<}=Z_{\psi}^{-1/2}\psi_{\xi} \ , ~\phi_<=Z_{\phi}^{-1/2}\phi \ .
\end{eqnarray*}
Note that the $z$- and transverse directions may have different scaling exponents on account of the 
breaking of the full rotation symmetry. The wavefunction renormalization constants $Z_{\psi}$ and 
$Z_{\phi}$ are chosen to bring the terms $\int_X\psi^{\dagger}_{\xi<}\partial_{\tau}\psi_{\xi<}$ and 
$i\! \int_X\phi_<\rho_{0<}$ in the action back to their original forms. In this way, we obtain an 
effective action for the slow modes, which has the same form as $S$ [Eq. (\ref{hodsms1})] but with 
renormalized parameters.

We further choose the values of $z$ and $z_3$ such that $v_3$ and $b$ are both RG invariants, yielding
\begin{equation}
 z=2-\beta^3\alpha G_4(\beta,\delta) \ , ~z_3=z+\beta\alpha G_3(\beta,\delta) \ , \label{hodsmrg11}
\end{equation}
where $r=v_{\perp}/v_3$ is the ratio of the components of the velocity, $\eta\equiv g_3^2/g_{\perp}^2$ 
is the anisotropy parameter, $\beta=\eta/r^2$, and $\delta=\tilde{b}\Lambda$. Therefore, the one-loop RG 
equations for these dimensionless parameters are given by
\begin{eqnarray}
 \frac{dr_l}{dl} \! \! &=& \! \! r_l+r_l\beta_l^2\alpha_l[G_{\perp}(\beta_l,\delta_l)-\beta_lG_4
 (\beta_l,\delta_l)] \ , \label{hodsmrg17} \\
 \frac{d\delta_l}{dl} \! \! &=& \! \! -3\delta_l-2\beta^2_l\alpha_l\delta_l[G_{\perp}(\beta_l,\delta_l)
 -\beta_lG_4(\beta_l,\delta_l)] \ , ~~\label{hodsmrg12} \\
 \frac{d\alpha_l}{dl} \! \! &=& \! \! -\alpha_l^2[\beta_lG_3(\beta_l,\delta_l)+2F_{\perp}(\delta_l)] 
 \ , \label{hodsmrg13} \\
 \frac{d\beta_l}{dl} \! \! &=& \! \! 2\beta_l^2\alpha_l[G_3(\beta_l,\delta_l)-\beta_lG_{\perp}
 (\beta_l,\delta_l)] \nonumber \\
 \! \! & & \! \! +2\beta_l\alpha_l[F_{\perp}(\delta_l)-\beta_lF_3(\delta_l)] \ , \label{hodsmrg14}
\end{eqnarray}
where $A_l$ is the value of $A$ at the scale $l$. In the above, 
\begin{eqnarray*}
 F_3(z)=\frac{2}{3(1+z^2)} \ , ~F_{\perp}(z)=\frac{2+11z^2+8z^4}{3(1+z^2)^2} \ ,
\end{eqnarray*}
and
\begin{eqnarray*}
 G_{\perp}(\beta,z) \! \! &=& \! \! \! \int^{+\infty}_0 \! \frac{dt}{(t^2+\beta)^2\sqrt{t^2+1+z^2}} \ , 
 \\
 G_3(\beta,z) \! \! &=& \! \! \! \int^{+\infty}_0 \! dt\frac{2t^2}{(t^2+\beta)^2\sqrt{t^2+1+z^2}} \ , \\
 G_4(\beta,z) \! \! &=& \! \! \! \int^{+\infty}_0 \! \frac{dt}{(t^2+\beta)^3\sqrt{t^2+1+z^2}} \ .
\end{eqnarray*}
The details of the derivation of Eqs. (\ref{hodsmrg17}) -- (\ref{hodsmrg14}) are left in appendix 
\ref{a2}.

\begin{figure}
	\begin{center}
		\includegraphics[width=0.99\columnwidth]{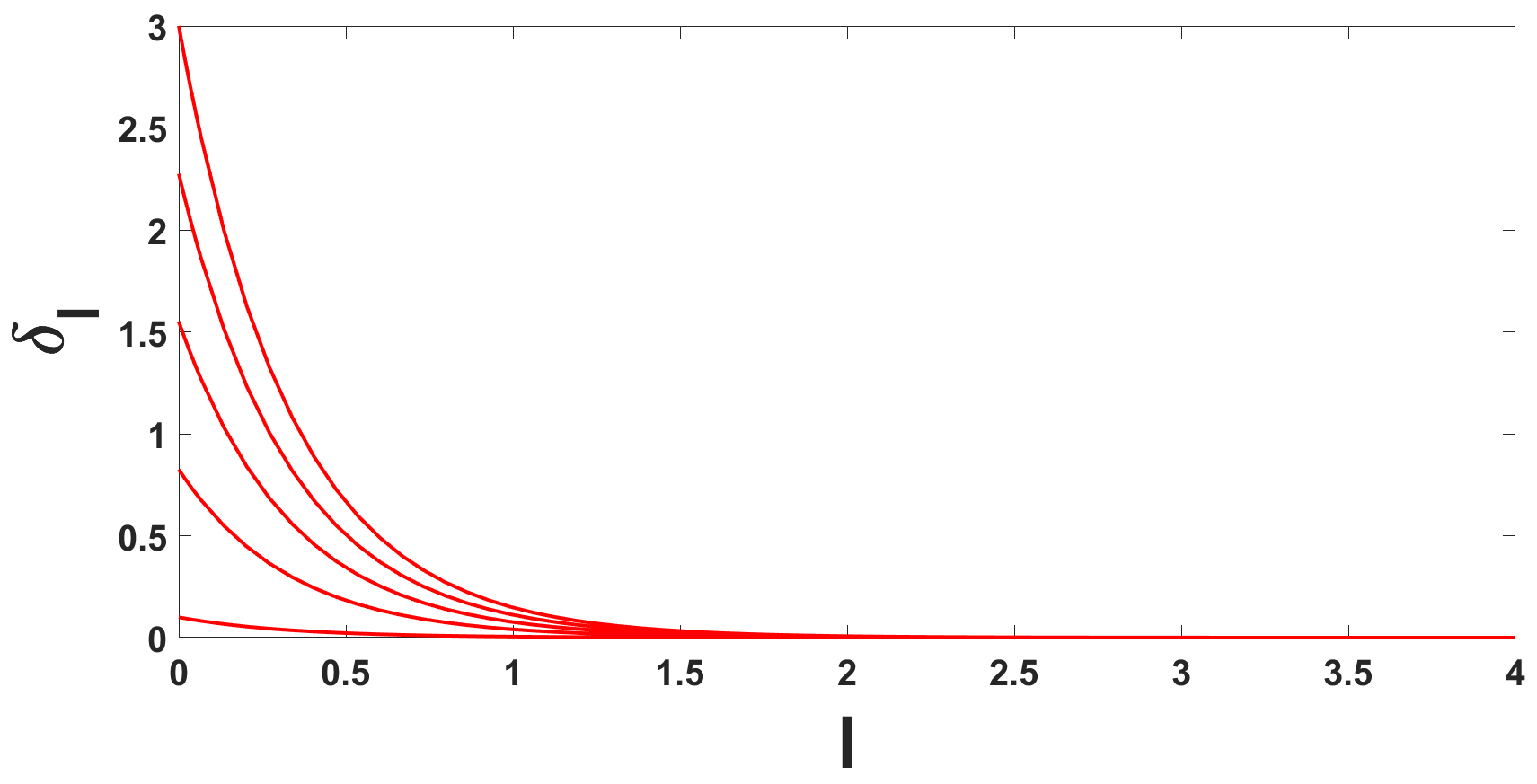}
		\caption{The RG flow of $\delta$ with $\alpha_0=0.1$ and $\beta_0=0.3$.}
		\label{hodsmrgf1} 
	\end{center}
\end{figure}

Equations (\ref{hodsmrg12}) -- (\ref{hodsmrg14}) themselves form a closed set under RG transformations. 
Because $F_3(0)=2/3=F_{\perp}(0)$ and $G_3(1,0)=2/3=G_{\perp}(1,0)$, this set of one-loop RG equations 
has two fixed points: $(\alpha,\beta,\delta)=(0,0,0)$ and $(0,1,0)$. Around the two fixed points, 
$\delta$ always flows to zero at low energies, and thus it is an irrelevant parameter in the sense of RG. 
(Figure \ref{hodsmrgf1} plots the RG flow of $\delta$.) 

\begin{figure}
	\begin{center}
		\includegraphics[width=0.99\columnwidth]{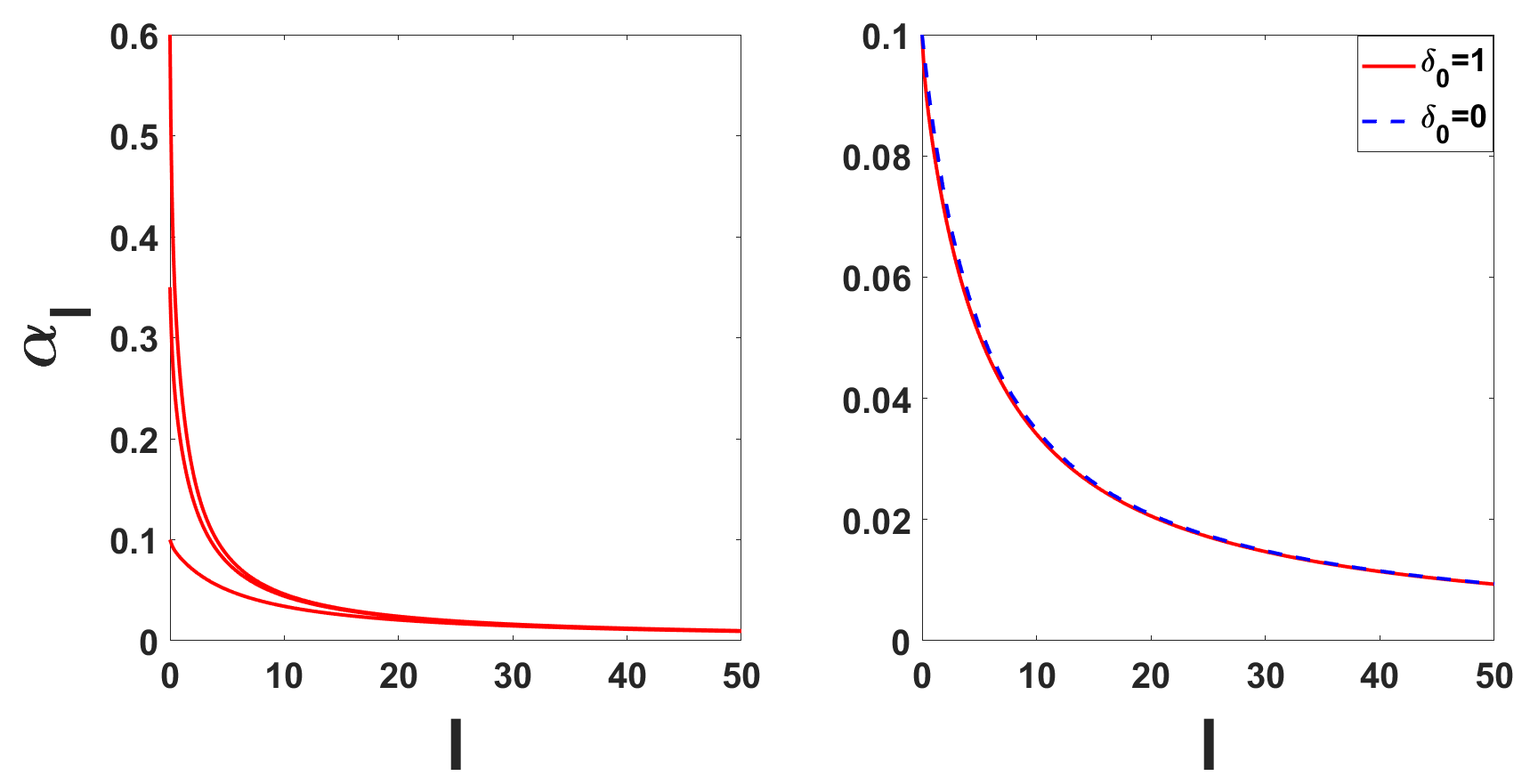}
		\includegraphics[width=0.99\columnwidth]{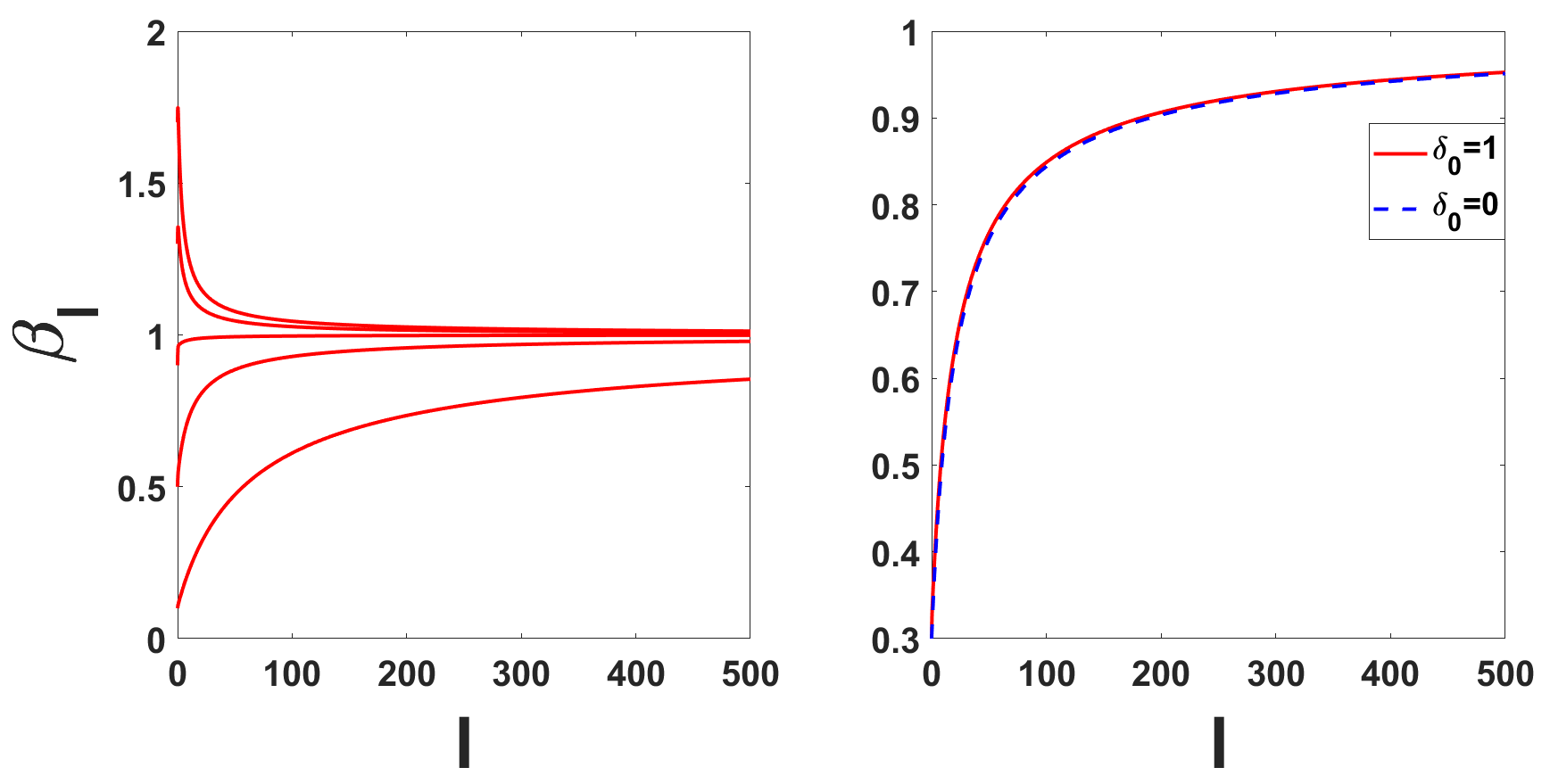}
		\caption{{\bf Top}: The RG flows of $\alpha$ with $\beta_0=0.3$, $\delta_0=1$, and different values of 
			$\alpha_0$ (left) and the RG flows of $\alpha$ with $\alpha_0=0.1$, $\beta_0=0.3$, and different values 
			of $\delta_0$ (right). {\bf Bottom}: The RG flows of $\beta$ with $\alpha_0=0.1$, $\delta_0=1$, and 
			different values of $\beta_0$ (left) and the RG flows of $\beta$ with $\alpha_0=0.1$, $\beta_0=0.3$, 
			and different values of $\delta_0$ (right). The curves for $\delta_0\neq 0$ and $\delta_0=0$ almost 
			coincide with each other due to the irrelevancy of $\delta$ under RG transformations.}
		\label{hodsmrgf11} 
	\end{center}
\end{figure}

The typical RG flows of $\alpha$ and $\beta$ are depicted in Fig. \ref{hodsmrgf11}. We see that $\alpha$ 
is marginally irrelevant and flows to zero at low energies. On the other hand, $\beta$ flows to $1$ at 
low energies. These facts indicate that the fixed point $(\alpha,\beta,\delta)=(0,0,0)$ is IR unstable 
while the fixed point $(\alpha,\beta,\delta)=(0,1,0)$ is IR stable. It is the latter which controls the 
low-energy physics of the second-order DSM in the presence of long range Coulomb repulsion. 

\begin{figure}
\begin{center}
 \includegraphics[width=0.99\columnwidth]{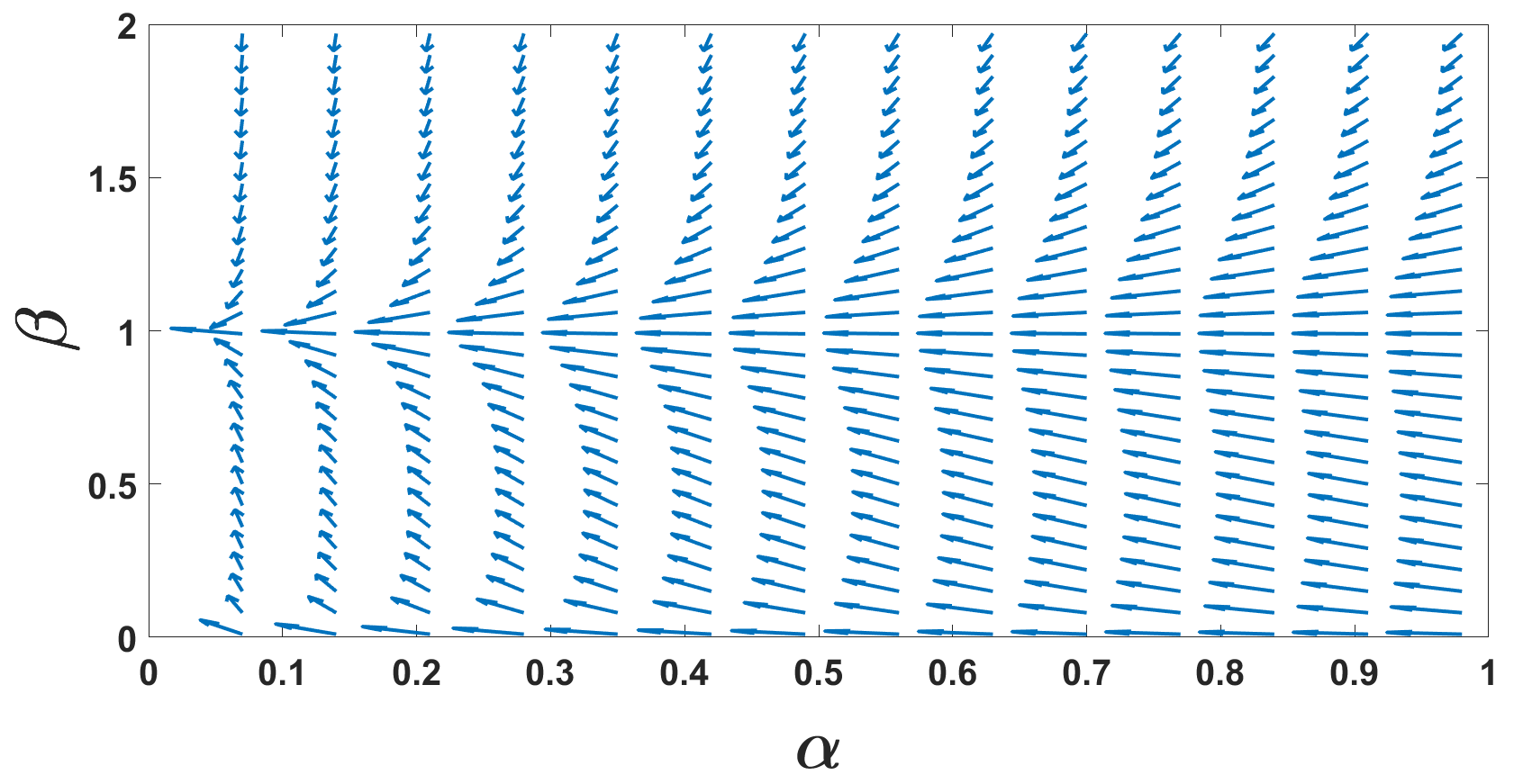}
 \caption{The flow diagram of $\alpha$ and $\beta$. It is clear that the RG flow converges to the line 
 $\beta=1$.}
 \label{hodsmrgf12} 
\end{center}
\end{figure}

Since $\delta$ flows to zero quickly at low energies, we may set $\delta_l=0$ in Eqs. (\ref{hodsmrg13}) 
and (\ref{hodsmrg14}), yielding
\begin{eqnarray}
 \frac{d\alpha_l}{dl} \! \! &=& \! \! -\alpha_l^2 \! \left[\beta_lG_3(\beta_l,0)+\frac{4}{3}\right] , 
 \label{hodsmrg15} \\
 \frac{d\beta_l}{dl} \! \! &=& \! \! 2\beta_l^2\alpha_l[G_3(\beta_l,0)-\beta_lG_{\perp}(\beta_l,0)] 
 \nonumber \\
 \! \! & & \! \! +\frac{4}{3}\beta_l\alpha_l(1-\beta_l) \ . \label{hodsmrg16}
\end{eqnarray}
Equations (\ref{hodsmrg15}) and (\ref{hodsmrg16}) correctly produce the RG flows of $\alpha$ and $\beta$,
as illustrated in Fig. \ref{hodsmrgf11}. We plot the flow diagram of $\alpha$ and $\beta$ in Fig. 
\ref{hodsmrgf12} according to Eqs. (\ref{hodsmrg15}) and (\ref{hodsmrg16}), which reveals the nature of 
the two fixed points $(\alpha,\beta,\delta)=(0,0,0)$ and $(0,1,0)$. We see that $\beta$ flows to $1$ 
quickly under the RG transformations, so that the RG flow converges to the line $\beta=1$. Along the 
surface $\beta_l=1$ and $\delta_l=0$, Eqs. (\ref{hodsmrg17}) and (\ref{hodsmrg15}) can be simplified and 
their solutions are
\begin{equation}
 r_l=r_0e^l(1+2\alpha_0l)^{1/15} \ , ~\alpha_l=\frac{\alpha_0}{1+2\alpha_0l} \ , \label{hodsmrgsol1}
\end{equation}
respectively.

\subsection{The specific heat of the interacting DSM}

We are able to calculate the specific heat for the interacting DSM with the help of the RG equations\cite{Jian,Sheehy}. 
Let $f_l=f(T_l,r_l,v_{3l},b_l,\alpha_l,\beta_l)$ be the free energy density at the scale $l$, which can 
be written as
\begin{eqnarray*}
 f_l=-\frac{T_l}{V_l}\ln{Z} \ ,
\end{eqnarray*}
where $Z$ is the partition function, and $T_l$, $V_l$ are determined by the equations
\begin{equation}
 \frac{dT_l}{dl}=zT_l \ , \label{hodsmcv11}
\end{equation}
and
\begin{eqnarray*}
 \frac{dV_l}{dl}=-(2+z_3)V_l \ ,
\end{eqnarray*}
with the initial conditions $T_0=T$ and $V_0=V$, respectively. Since we keep $Z$ to be invariant under 
RG transformations, $f_l$ satisfies the equation
\begin{eqnarray*}
 \frac{df_l}{dl}=(2+z_3+z)f_l \ .
\end{eqnarray*}
with the initial condition $f_0=f$ where $f$ is the free energy density of the system. We define $c_l$ 
as
\begin{eqnarray*}
 c_l=-T_l\frac{\partial^2f_l}{\partial T_l^2} \ .
\end{eqnarray*}
Then, $c_l$ obeys the equation
\begin{equation}
 \frac{dc_l}{dl}=(2+z_3)c_l \ , \label{hodsmcv12}
\end{equation}
with the initial condition $c_0=c_v$ where $c_v$ is the specific heat of the system. We will run the RG 
transformation to the scale $l_*$ such that $T_*=D$ where $A_*=A(l_*)$ and $D$ is of the order of the 
band width. 

According to the previous RG analysis, we have $v_{3*}=v_3$, $\beta_*\approx 1$, $b_*=b$, 
$\delta_*\approx 0$, and
\begin{eqnarray*}
 z=2-\frac{8}{15}\alpha_l \ , ~z_3=2+\frac{2}{15}\alpha_l \ .
\end{eqnarray*}
Inserting these into Eq. (\ref{hodsmcv11}) and using Eq. (\ref{hodsmrgsol1}), we get the solution of Eq.
(\ref{hodsmcv11})
\begin{eqnarray*}
 T_l=\frac{Te^{2l}}{(1+2\alpha_0l)^{4/15}} \ ,
\end{eqnarray*}
which results in
\begin{equation}
 \frac{D}{T}=\frac{e^{2l_*}}{(1+2\alpha_0l_*)^{4/15}} \ . \label{hodsmcv14}
\end{equation}
Equation (\ref{hodsmcv14}) determines $l_*$ as a function of $T/D$. 

In terms of these results, the solution to Eq. (\ref{hodsmcv12}) is
\begin{eqnarray*}
 c_l=c_ve^{4l}(1+2\alpha_0l)^{1/15} \ ,
\end{eqnarray*}
which leads to
\begin{eqnarray*}
 c_v=\frac{c_*e^{-4l_*}}{(1+2\alpha_0l_*)^{1/15}} \ .
\end{eqnarray*}
Here $c_*$ is approximately given by the specific heat of a non-interacting DSM at temperature $T_*=D$,
with $v_{\perp}$ replaced by $v_{\perp *}=v_3r_*$.

\begin{figure}
	\begin{center}
		\includegraphics[width=0.99\columnwidth]{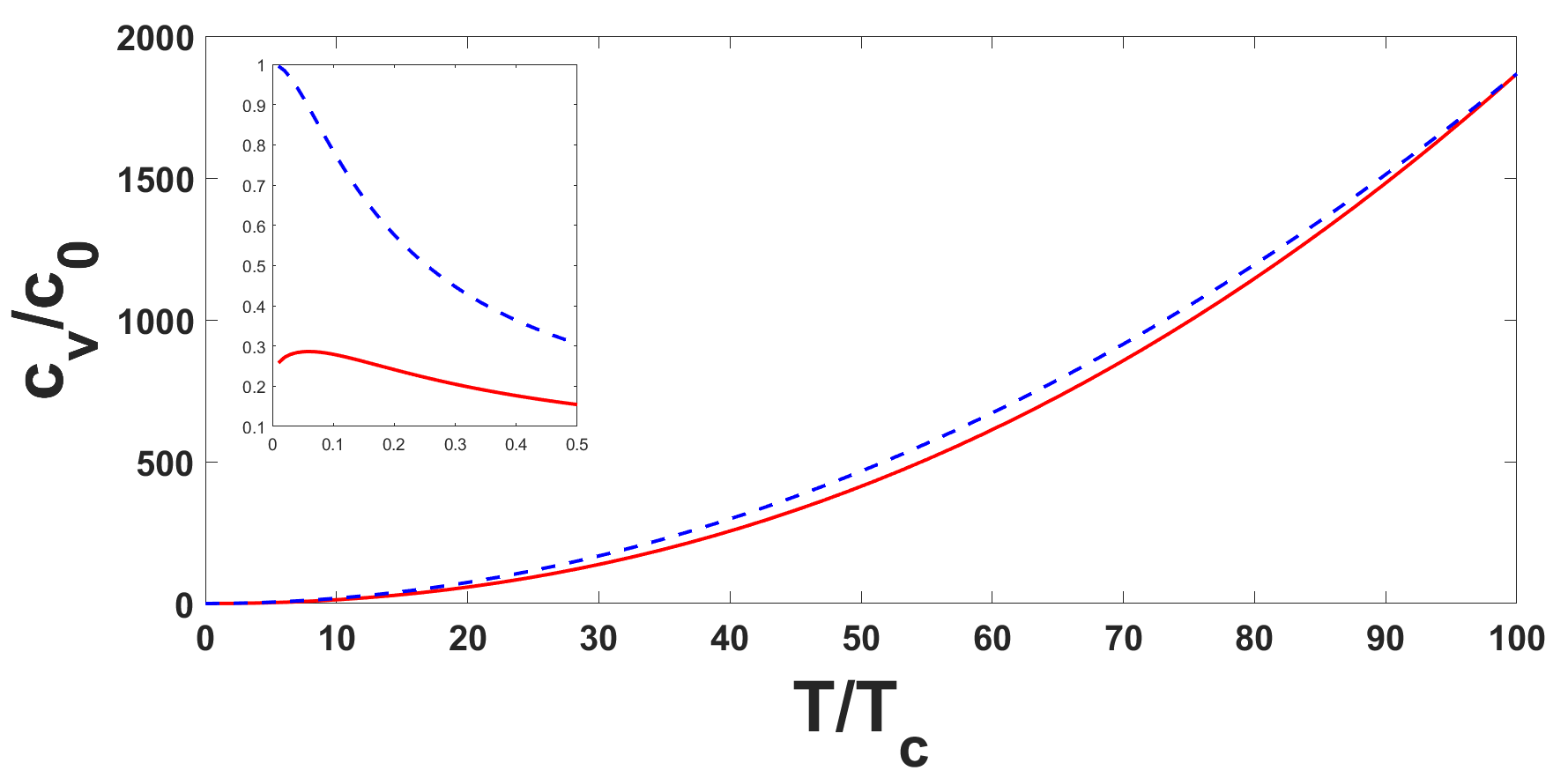}
		\caption{$c_v/c_0$ as a function of $T/T_c$ with $\alpha_0=0.3$ and $T_c/D=0.01$, where $c_0$ is the 
			specific heat of the first-order DSM at $T=T_c$. The solid and dashed lines correspond to the cases with
			and without the Coulomb interaction, respectively. The inset shows $c_v$, in units of the specific heat 
			for a non-interacting first-order DSM, for $T/T_c<0.5$.}
		\label{hodsmcvf11} 
	\end{center}
\end{figure}

With the help of Eqs. (\ref{hodsmcv1}) and (\ref{hodsmcv14}), we get $c_v$ for the second-order DSM
\begin{eqnarray}
 c_v(T) &=& \frac{T^3}{2v_{\perp}^2v_3\pi^2(1+2\alpha_0l_*)} \nonumber \\
 & & \times \! \int^{+\infty}_0 \! dx\frac{g(Tx/[T_cf(T)])x^4e^{-x}}{(1+e^{-x})^2} \ , \label{hodsmcv15}
\end{eqnarray}
where $f(T)=(1+2\alpha_0l_*)^{2/5}$. The temperature dependence of $c_v$ given by Eq. (\ref{hodsmcv15}) 
is shown in Fig. \ref{hodsmcvf11}, by assuming that $T_c\ll D$. We measure $c_v$ in units of $c_0$ where
$c_0=c_1(T_c)$ is the specific heat of the first-order DSM at $T=T_c$. 

A few points should be emphasized. (i) First of all, the value of $c_v$ is suppressed in the presence of 
the Coulomb interaction when $T\ll D$. However, it approaches the one for a non-interacting second-order 
DSM at high temperatures, as it should be. (ii) Next, as shown in the inset, even in the absence of the 
Coulomb interaction, the value of $c_v$ is close to the one for a non-interacting first-order DSM only 
at extremely low temperatures. In the presence of the Coulomb interaction, the deviation is more 
significant.

\section{Conclusions}

We have studied the effects of Coulomb interactions on a second-order DSM. In contrast with the 
first-order DSM, the full rotation symmetry is broken so that the low-energy physics is controlled by
two dimensionless parameters: the dimensionless coupling constant $\alpha$ and the ratio of anisotropy 
parameters $\beta$. In terms of the one-loop RG equations, we show that $\alpha$ is marginally irrelevant 
and $\beta$ flows to $1$ at low energies. Therefore, physical quantities at low temperatures can be 
computed in terms of a renormalized perturbative expansion in $\alpha$. As an application, we calculate
the specific heat in terms of the RG. We find that the presence of the Coulomb interaction suppresses the
value of specific heat, compared to the one for a non-interacting second-order DSM.

Another consequence following from the breaking of the full rotation symmetry is the existence of a 
crossover temperature $T_c$ and an associated length scale $L_c$. When $T\ll T_c$, the values of physical 
quantities will be close to those for the first-order DSM; otherwise significant deviations will be 
observed. Similar phenomena will occur in the spatial dependence of the charge density produced by a 
charged impurity. When the distance $r$ from the impurity is smaller than $L_c$, the induced charge 
density will be anisotropic. On the other hand, the tail of the induced charge density is isotropic.

In the present work, we derive the RG equation for the coupling constant within two different regimes: 
the large N limit [Eq. (\ref{hodsmrg25})] and the perturbative expansion to the one-loop order [Eq. 
(\ref{hodsmrg13})]. In both cases, the (dimensionless) coupling constant flows to zero at low energies.
Therefore, the stability of the second-order DSM in the presence of the Coulomb interaction may goes 
beyond the weak-coupling regime.

If we set $b=0$, the Hamiltonian of the second-order DSM will reduce to the one of the first-order DSM.
This corresponds to the limit $T_c\rightarrow+\infty$ or $L_c\rightarrow 0$. Our results become those 
for the first-order DSM in this limit. On the other hand, if we consider the highly anisotropic limit, 
i.e., $v_{\perp}=0$, the Hamiltonian of the second-order DSM will be similar to that of a double WSM, 
except that the $2\times 2$ Pauli matrices in the double WSM are replaced the $4\times 4$ $\Gamma$ 
matrices in the present case. This corresponds to the limit $T_c\rightarrow 0$ or $L_c\rightarrow +\infty$. 
In fact, the functional form of the vacuum polarization in this limit is identical to that for a double 
WSM\cite{Lai,Jian}.

Most of our results are based on the one-loop RG equations, which hold only in the weak coupling regime. 
At strong coupling, the DSM may become unstable toward other phases. For the first-order DSM, it has been 
shown that the phase diagram may depend on the number $N$ of the species of the Dirac fermions\cite{Nomura,Gonzalez}.
Based on a U($1$) lattice gauge theory with $N=4$, a Mott insulating phase is identified when the  
Coulomb interaction becomes strong\cite{Nomura}. By combining the RG method and the self-consistent 
resolution of Schwinger-Dyson equations, the strong Coulomb interaction results in dynamical mass 
generation for $N\leq 4$ and a line of critical points characterized by the suppression of the 
quasiparticle weight at low energies for $N\gg 1$\cite{Gonzalez}. It is interesting to see whether or not
similar phenomena will occur for the second-order DSM.

\acknowledgments

The works of Y.-W. Lee and Y.L. Lee are supported by the Ministry of Science and Technology, Taiwan,
under the grant number MOST 109-2112-M-029-004 and MOST 109-2112-M-018-006, respectively.

\appendix
\section{The vacuum polarization}
\label{a1}

Here we present the details of the calculation for the vacuum polarization. Performing the trace over the 
$\Gamma$ matrices, we find that
\begin{widetext}
\begin{eqnarray*}
 \Pi(0,0,q_3) \! \! &=& \! \! \frac{8}{v_{\perp}^2v_3} \! \int^{\prime} \! \! \frac{d^3\tilde{p}}
 {(2\pi)^3} \! \int^{+\infty}_{-\infty} \! \frac{dp_0}{2\pi}
 \frac{p_0^2-\tilde{p}_3(\tilde{p}_3+\tilde{q}_3)-\tilde{\bm{p}}_{\perp}^2-\tilde{b}^2\tilde{\bm{p}}_{\perp}^4}
 {(p_0^2+\tilde{p}_3^2+\tilde{\bm{p}}_{\perp}^2+\tilde{b}^2\tilde{\bm{p}}_{\perp}^4)
 [p_0^2+(\tilde{p}_3+\tilde{q}_3)^2+\tilde{\bm{p}}_{\perp}^2+\tilde{b}^2\tilde{\bm{p}}_{\perp}^4]} \\
 \! \! &=& \! \! \frac{8}{v_{\perp}^2v_3} \! \int^1_0 \! dx \! \int^{\prime} \! \! \frac{d^3\tilde{p}}
 {(2\pi)^3} \! \int^{+\infty}_{-\infty} \! \frac{dp_0}{2\pi} \frac{p_0^2-\tilde{p}_3(\tilde{p}_3+\tilde{q}_3)-\tilde{\bm{p}}_{\perp}^2-\tilde{b}^2\tilde{\bm{p}}_{\perp}^4}
 {[p_0^2+(\tilde{p}_3+\tilde{q}_3x)^2+\tilde{\bm{p}}_{\perp}^2+\tilde{b}^2\tilde{\bm{p}}_{\perp}^4+\tilde{q}_3^2x(1-x)]^2}
 \\
 \! \! &=& \! \! \frac{2\tilde{q}_3^2}{v_{\perp}^2v_3\pi^2} \! \int^1_0 \! dxx(1-x) \! \int^{\Lambda}_0 
 \! d\tilde{p}_{\perp}\frac{\tilde{p}_{\perp}}{\tilde{p}_{\perp}^2+\tilde{b}^2\tilde{p}_{\perp}^4+\tilde{q}_3^2x(1-x)} 
 \ ,
\end{eqnarray*}
and
\begin{eqnarray*}
 & & \! \! \! \Pi(0,\bm{q}_{\perp},0)=\frac{8}{v_{\perp}^2v_3} \! \int^{\prime} \! \! \frac{d^3\tilde{p}}
 {(2\pi)^3} \! \int^{+\infty}_{-\infty} \! \! \frac{dp_0}{2\pi}
 \frac{p_0^2-\tilde{p}_3^2-\tilde{\bm{p}}_{\perp}\cdot(\tilde{\bm{p}}_{\perp}+\tilde{\bm{q}}_{\perp})
 -\tilde{b}^2 \! \sum_{j=4,5}d_j(\tilde{\bm{p}}_{\perp})d_j(\tilde{\bm{p}}_{\perp}+\tilde{\bm{q}}_{\perp})}
 {(p_0^2+\tilde{p}_3^2+\tilde{\bm{p}}_{\perp}^2+\tilde{b}^2\tilde{\bm{p}}_{\perp}^4)
 [p_0^2+\tilde{p}_3^2+(\tilde{\bm{p}}_{\perp}+\tilde{\bm{q}}_{\perp})^2+\tilde{b}^2(\tilde{\bm{p}}_{\perp}+\tilde{\bm{q}}_{\perp})^4]} 
 \\
 & & \! \! \! =\frac{8}{v_{\perp}^2v_3} \! \int^1_0 \! \! dx \! \int^{\prime} \! \! \frac{d^3\tilde{p}}
 {(2\pi)^3} \! \int^{+\infty}_{-\infty} \! \frac{dp_0}{2\pi}\frac{p_0^2-\tilde{p}_3^2
 -\tilde{\bm{p}}_{\perp}\cdot(\tilde{\bm{p}}_{\perp}+\tilde{\bm{q}}_{\perp})
 -\tilde{b}^2 \! \sum_{j=4,5}d_j(\tilde{\bm{p}}_{\perp})d_j(\tilde{\bm{p}}_{\perp}+\tilde{\bm{q}}_{\perp})}
 {\{p_0^2+\tilde{p}_3^2+(1-x)(\tilde{\bm{p}}_{\perp}^2+\tilde{b}^2\tilde{\bm{p}}_{\perp}^4)
 +x[(\tilde{\bm{p}}_{\perp}+\tilde{\bm{q}}_{\perp})^2+\tilde{b}^2(\tilde{\bm{p}}_{\perp}+\tilde{\bm{q}}_{\perp})^4]\}^2} 
 \\
 & & \! \! \! =\frac{2}{v_{\perp}^2v_3\pi} \! \int^1_0 \! \! dx \! \int_{\Lambda} \! \! 
 \frac{d\tilde{p}_1d\tilde{p}_2}{(2\pi)^2} \! \left[1-\frac{\tilde{\bm{p}}_{\perp}\cdot(\tilde{\bm{p}}_{\perp}+\tilde{\bm{q}}_{\perp})
 +\tilde{b}^2(\tilde{p}_2^2-\tilde{p}_1^2)[(\tilde{p}_2+\tilde{q}_2)^2-(\tilde{p}_1+\tilde{q}_1)^2]
 +4\tilde{b}^2\tilde{p}_1\tilde{p}_2(\tilde{p}_1+\tilde{q}_1)(\tilde{p}_2+\tilde{q}_2)}
 {(1-x)(\tilde{\bm{p}}_{\perp}^2+\tilde{b}^2\tilde{\bm{p}}_{\perp}^4)
 +x[(\tilde{\bm{p}}_{\perp}+\tilde{\bm{q}}_{\perp})^2+\tilde{b}^2(\tilde{\bm{p}}_{\perp}+\tilde{\bm{q}}_{\perp})^4]}
 \right] ,
\end{eqnarray*}
where $\int^{\prime} \! \! d^3\tilde{p}=\! \int_{|\tilde{\bm{p}}_{\perp}|<\Lambda} \! d\tilde{p}_1d\tilde{p}_2 
\! \int^{+\infty}_{-\infty} \! d\tilde{p}_3$. In the above, we have used the identity
\begin{eqnarray*}
 \int^{+\infty}_{-\infty} \! dp_0\frac{p_0^2-\Delta^2}{(p_0^2+\Delta^2)^2}=0 \ ,
\end{eqnarray*}
and the dimensional regularization to perform the integration over $p_0$ and $\tilde{p}_3$. For the case 
with $\bm{q}_{\perp}=0$ and $q_3\neq 0$, we first perform the $x$ integral, yielding Eq. (\ref{hodsmvc22}).

For the case with $\bm{q}_{\perp}\neq 0$ and $q_3=0$, without loss of generality, we choose the 
coordinate frame such that $\tilde{\bm{q}}_{\perp}=(\tilde{q}_{\perp},0)$. Therefore, we have
\begin{eqnarray*}
 \frac{\partial}{\partial\Lambda}\Pi(0,\bm{q}_{\perp},0)=\frac{\Lambda}{2v_{\perp}^2v_3\pi^2}
 I_{\perp}(t,\tilde{b}\tilde{q}_{\perp}) \ ,
\end{eqnarray*}
where $t=\Lambda/\tilde{q}_{\perp}$ and
\begin{eqnarray*}
 I_{\perp}(t,s)=\! \int^1_0 \! \! dx \! \int^{2\pi}_0 \! \frac{d\phi}{\pi} \! \left[1
 -\frac{t^2+t\cos{\phi}+s^2t^2[\cos{(2\phi)}+2t\cos{\phi}+t^2]}
 {(1-x)(t^2+s^2t^4)+x[1+2t\cos{\phi}+t^2+s^2(1+2t\cos{\phi}+t^2)^2]}\right] ,
\end{eqnarray*}
For $|s|\ll 1$, we have
\begin{eqnarray*}
 I_{\perp}(t,s)\approx \! \int^1_0 \! \! dx \! \int^{2\pi}_0 \! \frac{d\phi}{\pi} \! \left(1
 -\frac{t+\cos{\phi}}{t+x/t+2x\cos{\phi}}\right) \! =2- \! \int^1_0 \! \! \frac{dx}{x} \! \left[1
 -\frac{(1-2t^2)x+t^2}{\sqrt{(1-4t^2)x^2+2t^2x+t^4}}\right] .
\end{eqnarray*}
Note that $t/x+1/t>2/x\geq 2$. When $t\gg 1$, this equation can be written as
\begin{eqnarray*}
 I_{\perp}(t,s)=2-2 \! \int^1_0 \! \! dx \! \left[1-\frac{2x(1-x)}{t^2}\right] \! +O(1/t^4)=\frac{2}
 {3t^2}+O(1/t^4) \ .
\end{eqnarray*}
Thus, we find that
\begin{eqnarray*}
 \Pi(0,\bm{q}_{\perp},0)\approx\frac{\tilde{q}_{\perp}^2}{3v_{\perp}^2v_3\pi^2} \! 
 \int^{\Lambda/\tilde{q}_{\perp}}_{\mu/\tilde{q}_{\perp}} \! \frac{dt}{t}+\cdots
 =\frac{\tilde{q}_{\perp}^2}{3v_{\perp}^2v_3\pi^2}\ln{\! \left(\frac{\Lambda}{\mu}\right)} \! +\cdots \ , 
\end{eqnarray*}
when $|\tilde{b}|\tilde{q}_{\perp}\ll 1$, where $\mu$ is an IR energy scale and $\cdots$ denotes the 
finite part.

On the other hand, for $|s|\gg 1$, we have
\begin{eqnarray*}
 I_{\perp}(t,s)\approx \! \int^1_0 \! \! dx \! \int^{2\pi}_0 \! \! \frac{d\phi}{\pi} \! \left[1
 -\frac{1+(2/t)\cos{\phi}+(1/t^2)\cos{(2\phi)}}
 {1+(4x/t)\cos{\phi}+(2x/t^2)(1+2\cos^2{\phi})+(4x/t^3)\cos{\phi}+x/t^4}\right] .
\end{eqnarray*}
\end{widetext}
When $t\gg 1$, we find that
\begin{eqnarray*}
 I_{\perp}(t,s) \! \! &\approx& \! \! \frac{2}{t^2} \! \int^1_0 \! \! dx \! \int^{2\pi}_0 \! \! 
 \frac{d\phi}{\pi}[x+(6x-8x^2)\cos^2{\phi}] \\
 \! \! & & \! \! +O(1/t^3) \\
 \! \! &=& \! \! \frac{8}{3t^2}+O(1/t^3) \ ,
\end{eqnarray*}
which leads to
\begin{eqnarray*}
 \Pi(0,\bm{q}_{\perp},0)\approx\frac{4\tilde{q}_{\perp}^2}{3v_{\perp}^2v_3\pi^2}
 \ln{\! \left(\frac{\Lambda}{\mu}\right)} \! +\cdots \ , 
\end{eqnarray*}
when $|\tilde{b}|\tilde{q}_{\perp}\gg 1$, where $\mu$ is an IR energy scale and $\cdots$ denotes the 
finite part.

These calculations suggest that the screened Coulomb potential $\mathcal{V}_s(\bm{q})$ in the momentum 
space takes the form
\begin{eqnarray*}
 \mathcal{V}_s(\bm{q})=\frac{g^2}{\bm{q}^2+w|q_3|} \ ,
\end{eqnarray*}
when $q\gg 1/L_c$. When $L_c$ is very large, the screened Coulomb potential in the real space is 
approximately given by
\begin{eqnarray*}
 V_s(\bm{r})\approx \! \int \! \! d^3qe^{i\bm{q}\cdot\bm{r}}\frac{g^2}{\bm{q}^2+w|q_3|} \ ,
\end{eqnarray*}
except its tail.

In the transverse direction, we have 
\begin{eqnarray*}
 & & \! \! V_s(\bm{r}_{\perp},0)=g^2 \! \int^{+\infty}_{-\infty} \! \frac{dp_3}{2\pi} \! \int \! \! 
 \frac{dp_1dp_2}{(2\pi)^2}\frac{e^{i\bm{p}_{\perp}\cdot\bm{r}_{\perp}}}{\bm{p}_{\perp}^2+p_3^2+w|p_3|} 
 \\
 & & \! \! =g^2 \! \int^{+\infty}_{-\infty} \! \frac{dp_3}{2\pi} \! \int^{+\infty}_0 \! \! 
 \frac{dp_{\perp}p_{\perp}}{(2\pi)^2} \! \int^{2\pi}_0 \! d\theta\frac{e^{ip_{\perp}r_{\perp}\cos{\theta}}}
 {p_{\perp}^2+p_3^2+w|p_3|} \ ,
\end{eqnarray*}
where $\bm{r}_{\perp}=(x,y)$ and $\bm{p}_{\perp}=(p_1,p_2)$. Using the Jacob-Auger expansion
\begin{eqnarray*}
 e^{ip_{\perp}r_{\perp}\cos{\theta}}=\! \sum_{m=-\infty}^{+\infty}i^me^{im\theta}J_m(p_{\perp}r_{\perp})
 \ ,
\end{eqnarray*}
where $J_m(x)$ is the Bessel function of order $m$, we find that
\begin{eqnarray*}
 V_s(\bm{r}_{\perp},0)=g^2 \! \int^{+\infty}_{-\infty} \! \! \frac{dp_3}{2\pi} \! \int^{+\infty}_0 \! \! 
 \frac{dp_{\perp}p_{\perp}}{(2\pi)^2}\frac{J_0(p_{\perp}r_{\perp})}{p_{\perp}^2+p_3^2+w|p_3|} \ .
\end{eqnarray*}
Integration over $p_{\perp}$ can be done with the help of the identity\cite{Grad}
\begin{eqnarray*}
 \int^{+\infty}_0 \! dx\frac{xJ_0(ax)}{x^2+k^2}=K_0(ak) \ ,
\end{eqnarray*}
where $a,k>0$ and $K_0(z)$ is the modified Bessel function of the second kind of order $0$, yielding
\begin{eqnarray*}
 V_s(\bm{r}_{\perp},0) \! \! &=& \! \! \frac{g^2}{4\pi^2} \! \int^{+\infty}_{-\infty} \! \frac{dp_3}
 {2\pi}K_0 \! \left(\sqrt{p_3^2+w|p_3|}r_{\perp}\right) \\
 \! \! &=& \! \! \frac{wg^2}{8\pi^3} \! \int^{+\infty}_0 \! dt\frac{t}{\sqrt{t^2+1}}K_0(wr_{\perp}t/2) 
 \\
 \! \! &=& \! \! \frac{g^2}{4\pi^3r_{\perp}} \! \int^{+\infty}_0 \! dt\frac{t}
 {\sqrt{t^2+(wr_{\perp}/2)^2}}K_0(t) \ .
\end{eqnarray*}
In terms of the identity\cite{Grad}
\begin{eqnarray*}
 \int^{+\infty}_0 \! dxx^{\mu}K_0(ax)=\frac{2^{\mu-1}}{a^{\mu+1}} \! \left[\Gamma \! \left(\frac{1+\mu}
 {2}\right)\right]^{\! 2} ,
\end{eqnarray*}
where $\mbox{Re}(a)>0$ and $\mbox{Re}(\mu+1)>0$, we get Eq. (\ref{hodsmvc11}).

On the other hand, in the $z$ direction, $V_s$ is given by
\begin{eqnarray*}
 V_s(0,z) \! \! &=& \! \! g^2 \! \int \! \! \frac{dp_1dp_2}{(2\pi)^2} \! \int^{+\infty}_{-\infty} \! 
 \frac{dp_3}{2\pi}\frac{e^{ip_3z}}{\bm{p}_{\perp}^2+p_3^2+w|p_3|} \\
 \! \! &=& \! \! g^2 \! \int \! \! \frac{dp_1dp_2}{(2\pi)^2} \! \int^{+\infty}_{-\infty} \! 
 \frac{dp_3}{2\pi} \frac{e^{ip_3|z|}}{\bm{p}_{\perp}^2+p_3^2+w|p_3|} \\
 \! \! &=& \! \! \frac{g^2}{|z|} \! \int \! \! \frac{dp_1dp_2}{(2\pi)^2} \! \int^{+\infty}_{-\infty} \! 
 \frac{dt}{2\pi}\frac{e^{it}}{\bm{p}_{\perp}^2+t^2z^2+w|z||t|}  \ .
\end{eqnarray*}
For $w|z|\ll 1$, $V_s(0,z)$ can be approximated as
\begin{eqnarray*}
 V_s(0,z) \! \! &\approx& \! \! \frac{g^2}{|z|} \! \int \! \! \frac{dp_1dp_2}{(2\pi)^2} \! 
 \int^{+\infty}_{-\infty} \! \frac{dt}{2\pi}\frac{e^{it}}{\bm{p}_{\perp}^2+t^2z^2} \\
 \! \! &=& \! \! g^2 \! \int \! \! \frac{dp_1dp_2}{(2\pi)^2} \! \int^{+\infty}_{-\infty} \! 
 \frac{dp_3}{2\pi}\frac{e^{ip_3|z|}}{\bm{p}_{\perp}^2+p_3^2} \\
 \! \! &=& \! \! \frac{g^2}{2} \! \int \! \! \frac{dp_1dp_2}{(2\pi)^2}\frac{e^{-p_{\perp}|z|}}{p_{\perp}} 
 =\frac{g^2}{4\pi|z|} \ .
\end{eqnarray*}

For $w|z|\gg 1$, $V_s(0,z)$ can be approximated as
\begin{eqnarray*}
 V_s(0,z) \! \! &\approx& \! \! \frac{g^2}{|z|} \! \int \! \! \frac{dp_1dp_2}{(2\pi)^2} \! 
 \int^{+\infty}_{-\infty} \! \frac{dt}{2\pi}\frac{e^{it}}{\bm{p}_{\perp}^2+w|z||t|} \\
 \! \! &=& \! \! g^2 \! \int \! \! \frac{dp_1dp_2}{(2\pi)^2} \! \int^{+\infty}_{-\infty} \! 
 \frac{dp_3}{2\pi}\frac{e^{ip_3z}}{\bm{p}_{\perp}^2+w|p_3|} \\
 \! \! &=& \! \! \frac{g^2}{\pi} \! \int \! \! \frac{dp_1dp_2}{(2\pi)^2} \! \int^{+\infty}_0 \! dp_3
 \frac{\cos{(p_3|z|)}}{\bm{p}_{\perp}^2+wp_3} \ .
\end{eqnarray*}
The $p_3$ integral can be performed with the help of the identity\cite{Grad}
\begin{eqnarray*}
 \int^{+\infty}_0 \! dx\frac{\cos{(ax)}}{x+\beta}=-\sin{(a\beta)}\mbox{si}(a\beta)-\cos{(a\beta)}
 \mbox{ci}(a\beta) \ ,
\end{eqnarray*}
where $a>0$, $|\mbox{arg}(\beta)|<\pi$, and $\mbox{si}(x)$, $\mbox{ci}(x)$ are respectively the sine and 
cosine integrals, yielding
\begin{eqnarray*}
 V_s(0,z) \! \! &=& \! \! -\frac{g^2}{4\pi^3w} \! \int \! \! dp_1dp_2\sin{(zp_{\perp}^2/w)}\mbox{si}
 (zp_{\perp}^2/w) \\
 \! \! & & \! \! -\frac{g^2}{4\pi^3w} \! \int \! \! dp_1dp_2\cos{(zp_{\perp}^2/w)}\mbox{ci}
 (zp_{\perp}^2/w) \\
 \! \! &=& \! \! -\frac{g^2}{4\pi^2w} \! \int^{+\infty}_0 \! dt\sin{(zt/w)}\mbox{si}(zt/w) \\
 \! \! & & \! \! -\frac{g^2}{4\pi^2w} \! \int^{+\infty}_0 \! dt\cos{(zt/w)}\mbox{ci}(zt/w) \ .
\end{eqnarray*}
Note that $\mbox{si}(-x)=-\mbox{si}(x)$ and $\mbox{ci}(-x)=\mbox{ci}(x)$. With the help of the identities\cite{Grad}
\begin{eqnarray*}
 & & \! \int^{+\infty}_0 \! \! dx\sin{(px)}\mbox{si}(qx)=-\frac{\pi}{4p} \\
 & & =\! \int^{+\infty}_0 \! \! dx\cos{(px)}\mbox{ci}(qx) \ ,
\end{eqnarray*}
when $p^2=q^2$, we find that
\begin{eqnarray*}
 V_s(0,z)=\frac{g^2}{8\pi|z|} \ . 
\end{eqnarray*}

\section{One-loop RG equations}
\label{a2}

By decomposing the fields into slow and fast modes, the action $S[\psi,\psi^{\dagger},\phi]$ can be 
written as
\begin{eqnarray*}
 S[\psi,\psi^{\dagger},\phi]=S[\psi_<,\psi^{\dagger}_<,\phi_<]+S[\psi_>,\psi^{\dagger}_>,\phi_>]
 +S_{int} \ ,
\end{eqnarray*}
where
\begin{eqnarray*}
 S_{int} \! =i \! \sum_{\xi} \! \int_X \! \! \left[\phi_>(\psi^{\dagger}_{\xi>}\psi_{\xi<}
 +\psi^{\dagger}_{\xi<}\psi_{\xi >})+\phi_<\psi^{\dagger}_{\xi>}\psi_{\xi>}\right] ,
\end{eqnarray*}
describes the coupling between the slow and fast modes. Note that the terms with a single fast mode 
vanish due to momentum conservation. By integrating out the fast modes, we get an effective action for 
the slow modes
\begin{eqnarray*}
 S_{eff}[\psi_<,\psi^{\dagger}_<,\phi_<]=S[\psi_<,\psi^{\dagger}_<,\phi_<]
 +I[\psi_<,\psi^{\dagger}_<,\phi_<] \ ,
\end{eqnarray*}
where
\begin{eqnarray*}
 & & e^{-I[\psi_<,\psi^{\dagger}_<,\phi_<]} \\
 & & =\! \int \! \! D[\psi_>]D[\psi^{\dagger}_>]D[\phi_>]e^{-S[\psi_>,\psi^{\dagger}_>,\phi_>]-S_{int}} 
 \ .
\end{eqnarray*}
In the following, we will compute $I$ in terms of a perturbative expansion in powers of $S_{int}$.

Before plunging into the detailed calculations, we notice that the action $S$ is invariant against the 
U($1$) gauge transformation
\begin{equation}
 \psi_{\xi}\rightarrow e^{-i\chi(\tau)}\psi_{\xi} \ , ~~\phi\rightarrow\phi+\partial_{\tau}\chi \ ,
 \label{hodsmwt1}
\end{equation}
where $\chi$ is a function of the imaginary time $\tau$. This gauge invariance results in a Ward identity, 
which puts a constraint on the low-energy physics. We now derive it.

By integrating out the fast modes, the action becomes 
\begin{eqnarray*}
 S \! \! &\rightarrow& \! \! \! \sum_{\xi=\pm} \! \int_X \! \! \psi^{\dagger}_{\xi}[(1+\Sigma_{\tau})
 \partial_{\tau}+\bar{h}_{\xi}]\psi_{\xi}+(1+\Gamma_v)i \! \int_X \! \! \phi\rho_0 \\
 \! \! & & \! \! +\frac{1}{2} \! \int_X \! \left[\! \left(\frac{1}{g_{\perp}^2}+\Pi_{\perp}\right) \!
 |\bm{\nabla}_{\perp}\phi|^2+ \! \left(\frac{1}{g_3^2}+\Pi_3\right) \! |\partial_3\phi|^2\right] \\
 \! \! & & \! \! +\cdots \ ,
\end{eqnarray*}
where $\bar{h}_{\xi}$ is the renormalized Hamiltonian whose actual form is irrelevant to our discussion
and $\cdots$ denotes the terms with higher scaling dimensions. The gauge invariance of $S$ requires that
\begin{equation}
 \Sigma_{\tau}=\Gamma_v \ . \label{hodsmwt11}
\end{equation}
Equation (\ref{hodsmwt11}) is the desired expression of the Ward identity.

To proceed, we write $I$ as
\begin{eqnarray*}
 I=\! \sum_{n=1}^{+\infty}I_n \ ,
\end{eqnarray*}
where $I_n$ denotes the contribution from $S_{int}^n$. We first compute $I_1$ which is given by
\begin{eqnarray*}
 I_1=\langle S_{int}\rangle_>=i \! \sum_{\xi} \! \int_X\phi_<\langle\psi^{\dagger}_{\xi>}\psi_{\xi>}
 \rangle_> \ ,
\end{eqnarray*}
where $\langle\cdots\rangle_>$ denotes the functional integral over the fast modes. Since
\begin{eqnarray*}
 & & \! \! \! \sum_{\xi}\langle\psi^{\dagger}_{\xi>}(X)\psi_{\xi>}(X)\rangle_> \\
 & & \! \! =\! \sum_{\xi} \! \int^{\prime} \! \frac{d^3p}{(2\pi)^3} \! \int^{+\infty}_{-\infty} \! 
 \frac{dp_0}{2\pi}e^{ip_00^+}\mbox{tr}[G_{\xi 0}(P)] \\
 & & \! \! =-\frac{8}{v_{\perp}^2v_3} \! \int^{\prime} \! \frac{d^3\tilde{p}}{(2\pi)^3} \! 
 \int^{+\infty}_{-\infty} \! \frac{dp_0}{2\pi}\frac{ip_0}
 {p_0^2+\tilde{\bm{p}}_{\perp}^2+\tilde{p}_3^2+\tilde{b}^2\tilde{\bm{p}}_{\perp}^4} \\
 & & \! \! =0 \ ,
\end{eqnarray*}
we conclude that $I_1=0$.

Next, we calculate $I_2$ which is given by
\begin{eqnarray*}
 I_2=-\frac{1}{2}\langle S_{int}^2\rangle_>+\frac{1}{2}I_1^2=-\frac{1}{2}\langle S_{int}^2\rangle_> \ .
\end{eqnarray*}
In view of $S_{int}$, $I_2$ consists of two terms:
\begin{eqnarray*}
 I_2 &=& \! \sum_{\xi} \! \int_X \! \int_Y\psi^{\dagger}_{\xi<}(X)\Sigma_{\xi}(X-Y)\psi_{\xi<}(Y) \\
 & & +\frac{1}{2} \! \int_X \! \int_Y\phi_<(X)\Pi(X-Y)\phi_<(Y) \ ,
\end{eqnarray*}
where $\Sigma_{\xi}$ and $\Pi$ are the self-energies of $\psi_{\xi<}$ and $\phi_<$, respectively. 

We first calculate the self-energy $\Pi$. In the momentum space, the one-loop contribution to it is of 
the form
\begin{eqnarray*}
 \Pi(Q)= \! \sum_{\xi} \! \int^{\prime} \! \! \frac{d^3p}{(2\pi)^3} \! \int^{+\infty}_{-\infty} 
 \! \! \frac{dp_0}{2\pi}\mbox{tr} \! \left[G_{\xi 0}(P)G_{\xi 0}(P+Q)\right] .
\end{eqnarray*}
By taking the trace over the $\Gamma$ matrices, $\Pi$ can be written as
\begin{widetext}
\begin{eqnarray*}
 \Pi(Q) \! \! &=& \! \! \frac{8}{v_{\perp}^2v_3} \! \int^{\prime} \! \frac{d^3\tilde{p}}{(2\pi)^3} \! 
 \int^{+\infty}_{-\infty} \! \frac{dp_0}{2\pi}
 \frac{p_0(p_0+q_0)-\sum_{j=1}^3\tilde{p}_j(\tilde{p}_j+\tilde{q}_j)-\tilde{b}^2 \! \sum_{j=4,5}
 d_j(\tilde{\bm{p}}_{\perp})d_j(\tilde{\bm{p}}_{\perp}+\tilde{\bm{q}}_{\perp})}
 {(p_0^2+[E(\tilde{\bm{p}})]^2)\{(p_0+q_0)^2+[E(\tilde{\bm{p}}+\tilde{\bm{q}})]^2\}} \\
 \! \! &=& \! \! \frac{4}{v_{\perp}^2v_3} \! \int^{\prime} \! \frac{d^3\tilde{p}}{(2\pi)^3}
 \frac{E(\tilde{\bm{p}}+\tilde{\bm{q}})[E(\tilde{\bm{p}}+\tilde{\bm{q}})+iq_0]+\sum_{j=1}^3\tilde{p}_j
 (\tilde{p}_j+\tilde{q}_j)+\tilde{b}^2 \! \sum_{j=4,5}d_j(\tilde{\bm{p}}_{\perp})d_j
 (\tilde{\bm{p}}_{\perp}+\tilde{\bm{q}}_{\perp})}{E(\tilde{\bm{p}}+\tilde{\bm{q}})
 [E(\tilde{\bm{p}}+\tilde{\bm{q}})-E(\tilde{\bm{p}})+iq_0]
 [E(\tilde{\bm{p}}+\tilde{\bm{q}})+E(\tilde{\bm{p}})+iq_0]} \\
 \! \! & & \! \! -\frac{4}{v_{\perp}^2v_3} \! \int^{\prime} \! \frac{d^3\tilde{p}}{(2\pi)^3}
 \frac{E(\tilde{\bm{p}})[E(\tilde{\bm{p}})-iq_0]+\sum_{j=1}^3\tilde{p}_j(\tilde{p}_j+\tilde{q}_j)
 +\tilde{b}^2 \! \sum_{j=4,5}d_j(\tilde{\bm{p}}_{\perp})d_j(\tilde{\bm{p}}_{\perp}+\tilde{\bm{q}}_{\perp})}
 {E(\tilde{\bm{p}})[E(\tilde{\bm{p}}+\tilde{\bm{q}})-E(\tilde{\bm{p}})+iq_0]
 [E(\tilde{\bm{p}}+\tilde{\bm{q}})+E(\tilde{\bm{p}})-iq_0]} \ ,
\end{eqnarray*}
\end{widetext}
where $E(\tilde{\bm{p}})=\sqrt{\tilde{\bm{p}}_{\perp}^2+\tilde{p}_3^2+\tilde{b}^2\tilde{\bm{p}}_{\perp}^4}$. 
From the first equality, we notice that $\Pi$ is an even function of $q_0$, $\tilde{q}_1$, $\tilde{q}_2$, 
and $\tilde{q}_3$. Moreover, it is invariant against the exchange of variables: 
$\tilde{q}_1\leftrightarrow\tilde{q}_2$. Therefore, the derivative expansion of $\Pi$ is of the form
\begin{eqnarray*}
 \Pi(Q)=\Pi(0)+\Pi_0q_0^2+\Pi_{\perp}\tilde{\bm{q}}_{\perp}^2+\Pi_3\tilde{q}_3^2+\cdots \ ,
\end{eqnarray*}
where $\cdots$ denotes the higher order terms. One may verify that 
$\lim_{q_0\rightarrow 0}\lim_{\bm{q}\rightarrow 0}\Pi(Q)=0=\lim_{\bm{q}\rightarrow 0}\lim_{q_0\rightarrow 0}\Pi(Q)$,
and thus $\Pi(0)=0$. This implies that the $\phi$ field is gapless to the one-loop order. 

The expansion coefficients $\Pi_3$ and $\Pi_{\perp}$ are given by
\begin{eqnarray*}
 \Pi_3 \! \! &=& \! \! \frac{1}{v_{\perp}^2v_3} \! \int^{\prime} \! \frac{d^3\tilde{p}}{(2\pi)^3} \! 
 \left\{\frac{1}{[E(\tilde{\bm{p}})]^3}-\frac{\tilde{p}_3^2}{[E(\tilde{\bm{p}})]^5}\right\} \\
 \! \! &=& \! \! \frac{l}{2v_{\perp}^2v_3\pi^2}F_3(\delta)+O(l^2) \ ,
\end{eqnarray*}
and
\begin{eqnarray*}
 \Pi_{\perp} \! \! &=& \! \! \frac{1}{v_{\perp}^2v_3} \! \int^{\prime} \! \frac{d^3\tilde{p}}{(2\pi)^3} 
 \! \left\{\frac{1+4\tilde{b}^2\bm{p}_{\perp}^2}{[E(\tilde{\bm{p}})]^3}
 -\frac{(1+2\tilde{b}^2\bm{p}_{\perp}^2)^2\bm{p}_{\perp}^2}{2[E(\tilde{\bm{p}})]^5}\right\} \\
 \! \! &=& \! \! \frac{l}{2v_{\perp}^2v_3\pi^2}F_{\perp}(\delta)+O(l^2) \ ,
\end{eqnarray*}
respectively, where $\delta=\tilde{b}\Lambda$ is a dimensionless parameter and 
\begin{eqnarray*}
 F_3(z) \! \! &=& \! \!\! \int^{+\infty}_0 \! \! dt \! \left[\frac{1}{(t^2+1+z^2)^{3/2}}-\frac{t^2}
 {(t^2+1+z^2)^{5/2}}\right] \\
 \! \! &=& \! \! \frac{2}{3(1+z^2)} \ ,
\end{eqnarray*}
\begin{eqnarray*}
 F_{\perp}(z) \! \! &=& \! \! \! \int^{+\infty}_0 \! \! dt \! \left[\frac{1+4z^2}{(t^2+1+z^2)^{3/2}}
 -\frac{(1+2z^2)^2}{2(t^2+1+z^2)^{5/2}}\right] \\
 \! \! &=& \! \! \frac{2+11z^2+8z^4}{3(1+z^2)^2} \ .
\end{eqnarray*}

Next, we calculate the self-energy $\Sigma_{\xi}$ of the fermionic fields. The one-loop contribution to 
it is of the form
\begin{eqnarray*}
 & & \! \! \Sigma_{\xi}(Q) \\
 & & \! \! =-\frac{i^2}{2}\cdot 2 \! \int^{\prime} \! \frac{d^3p}{(2\pi)^3} \! \int^{+\infty}_{-\infty} 
 \! \frac{dp_0}{2\pi}G_{0\xi}(P)D_0(\bm{q}-\bm{p}) \\
 & & \! \! =\! \int^{\prime} \! \frac{d^3p}{(2\pi)^3} \! \int^{+\infty}_{-\infty} \! \frac{dp_0}{2\pi}
 G_{0\xi}(P)D_0(\bm{q}-\bm{p}) \\
 & & \! \! =\Sigma_{\xi 0}(Q)+ \! \sum_{j=1}^5\Sigma_{\xi j}(Q)\Gamma_j \ ,
\end{eqnarray*}
where
\begin{eqnarray*}
 D_0(\bm{p})=-\frac{g_{\perp}^2g_3^2}{(g_3/v_{\perp})^2\tilde{\bm{p}}_{\perp}^2+(g_{\perp}/v_3)^2\tilde{p}_3^2}
 \ ,
\end{eqnarray*}
is the free propagator of the $\phi$ field,
\begin{eqnarray*}
 & & \! \! \Sigma_{\xi 0}(Q)=\frac{1}{4}\mbox{tr}[\Sigma_{\xi}(Q)] \\
 & & \! \! =-\! \int^{\prime} \! \frac{d^3p}{(2\pi)^3}D_0(\bm{q}-\bm{p}) \! \int^{+\infty}_{-\infty} \! 
 \frac{dp_0}{2\pi}\frac{ip_0}{p_0^2+[E(\tilde{\bm{p}})]^2} \\
 & & \! \! =0 \ ,
\end{eqnarray*}
and
\begin{eqnarray*}
 \Sigma_{\xi j}(Q)=\frac{1}{4}\mbox{tr}[\Gamma_j\Sigma_{\xi}(Q)] \ .
\end{eqnarray*}
Consequently, we find that
\begin{eqnarray*}
 & & \! \! \Sigma_{\xi j}(Q) \\
 & & \! \! =-\! \int^{\prime} \! \frac{d^3p}{(2\pi)^3}D_0(\bm{q}-\bm{p}) \! \int^{+\infty}_{-\infty} \! 
 \frac{dp_0}{2\pi}\frac{\tilde{p}_j}{p_0^2+[E(\tilde{\bm{p}})]^2} \\
 & & \! \! =-\! \int^{\prime} \! \frac{d^3p}{(2\pi)^3}\frac{\tilde{p}_jD_0(\bm{q}-\bm{p})}
 {2E(\tilde{\bm{p}})} \ ,
\end{eqnarray*}
for $j=1,2$,
\begin{eqnarray*}
 & & \! \! \Sigma_{\xi 3}(Q) \\
 & & \! \! =-\! \int^{\prime} \! \frac{d^3p}{(2\pi)^3}D_0(\bm{q}-\bm{p}) \! \int^{+\infty}_{-\infty} \! 
 \frac{dp_0}{2\pi}\frac{\xi\tilde{p}_3}{p_0^2+[E(\tilde{\bm{p}})]^2} \\
 & & \! \! =-\! \int^{\prime} \! \frac{d^3p}{(2\pi)^3}\frac{\xi\tilde{p}_3D_0(\bm{q}-\bm{p})}
 {2E(\tilde{\bm{p}})} \ ,
\end{eqnarray*}
and
\begin{eqnarray*}
 & & \! \! \Sigma_{\xi j}(Q) \\
 & & \! \! =-\! \int^{\prime} \! \frac{d^3p}{(2\pi)^3}D_0(\bm{q}-\bm{p}) \! \int^{+\infty}_{-\infty} \! 
 \frac{dp_0}{2\pi}\frac{\tilde{b}d_j(\bm{p}_{\perp})}{p_0^2+[E(\tilde{\bm{p}})]^2} \\
 & & \! \! =-\! \int^{\prime} \! \frac{d^3p}{(2\pi)^3}\frac{\tilde{b}d_j(\bm{p}_{\perp})D_0(\bm{q}-\bm{p})}
 {2E(\tilde{\bm{p}})} \ ,
\end{eqnarray*}
for $j=4,5$. 

To proceed, we perform the derivation expansions on $\Sigma_{\xi j}(Q)$. We notice that these functions 
have the following properties: (i) For $j=1,\cdots,5$, $\Sigma_{\xi j}(0)=0$. This can be seen as follows.
For $j=1,2,3$, 
\begin{eqnarray*}
 \Sigma_{\xi j}(0)\propto \! \int^{\prime} \! \frac{d^3p}{(2\pi)^3}\frac{\tilde{p}_jD_0(\bm{p})}
 {2E(\tilde{\bm{p}})}=0 \ ,
\end{eqnarray*}
because the integrand is an odd function of $p_j$. Next, for $j=4$,
\begin{eqnarray*}
 \Sigma_{\xi 4}(0)\propto \! \int^{\prime} \! \frac{d^3p}{(2\pi)^3}
 \frac{(\tilde{p}_2^2-\tilde{p}_1^2)D_0(\bm{p})}{2E(\tilde{\bm{p}})}=0 \ ,
\end{eqnarray*}
because the integrand is odd under the exchange of variables $\tilde{p}_1\leftrightarrow\tilde{p}_2$. 
Finally, for $j=5$,
\begin{eqnarray*}
 \Sigma_{\xi 5}(0)\propto \! \int^{\prime} \! \frac{d^3p}{(2\pi)^3}
 \frac{\tilde{p}_1\tilde{p}_2D_0(\bm{p})}{2E(\tilde{\bm{p}})}=0 \ ,
\end{eqnarray*}
because the integrand is an odd function of $p_1$ and $p_2$. The fact $\Sigma_{\xi j}(0)=0$ implies that 
the fermion fields are still gapless to the one-loop order. (ii) For $j=1,2,3$, $\Sigma_{\xi j}(Q)$ are 
odd functions of $q_j$ and even functions of $q_{i\neq j}$. (iii) $\Sigma_{\xi 4}(Q)$ is an even function 
of $q_1$, $q_2$, and $q_3$. Moreover, it is odd under the exchange of variables $q_1\leftrightarrow q_2$. 
(iv) $\Sigma_{\xi 5}(Q)$ is an odd function of $q_1$ and $q_2$ and an even function of $q_3$. Moreover, 
it is an even function of $\tilde{\bm{q}}_{\perp}$. Based on (i) -- (iv), the derivative expansions of 
$\Sigma_{\xi j}(Q)$ are of the forms:
\begin{eqnarray*}
 \Sigma_{\xi j}(Q)=\Sigma^{(1)}_{\xi j}\tilde{q}_j+\cdots \ ,
\end{eqnarray*}
for $j=1,2,3$ and
\begin{eqnarray*}
 \Sigma_{\xi j}(Q)=\Sigma^{(1)}_{\xi j}\tilde{b}d_j(\tilde{\bm{q}}_{\perp})+\Sigma^{(2)}_{\xi j}
 \tilde{q}_3^2+\cdots \ ,
\end{eqnarray*}
for $j=4,5$ where $\cdots$ denotes the higher order terms.

If $\Sigma^{(2)}_{\xi 4/5}\neq 0$, then we have to add additional terms to the bare action and introduce 
more parameters. Moreover, if $\Sigma^{(1)}_{\xi 4}\neq\Sigma^{(1)}_{\xi 5}$, we have to introduce two 
parameters $b_1$ and $b_2$ instead of a single one $b$. We will see, however, $\Sigma^{(2)}_{\xi 4/5}=0$ 
and $\Sigma^{(1)}_{\xi 4}=\Sigma^{(1)}_{\xi 5}$ to the one-loop order, and thus all terms listed in the 
bare action form a complete set under the RG transformations, at least to the one-loop order.

These expansions coefficients can be calculated as follows.
\begin{eqnarray*}
 \Sigma^{(1)}_{\xi 1/2} \! \! &=& \! \! -\frac{g_3^2}{v_{\perp}^2} \! \int^{\prime} \! \! \frac{d^3p}
 {(2\pi)^3}\frac{\tilde{p}_{1/2}^2D_0(\bm{p})}
 {E(\tilde{\bm{p}})[(g_3/v_{\perp})^2\tilde{\bm{p}}_{\perp}^2+(g_{\perp}/v_3)^2\tilde{p}_3^2]} \\
 \! \! &=& \! \! -\frac{g_3^2}{v^4_{\perp}v_3} \! \int^{\prime} \! \! \frac{d^3\tilde{p}}{(2\pi)^3}
 \frac{\tilde{p}_{1/2}^2D_0(\bm{p})}
 {E(\tilde{\bm{p}})[(g_3/v_{\perp})^2\tilde{\bm{p}}_{\perp}^2+(g_{\perp}/v_3)^2\tilde{p}_3^2]} \\
 \! \! &=& \! \! \beta^2\alpha l G_{\perp}(\beta,\delta)+O(l^2) \ , 
\end{eqnarray*}
\begin{eqnarray*}
 \Sigma^{(1)}_{\xi 3} \! \! &=& \! \! -\frac{\xi g_{\perp}^2}{v_3^2} \! \int^{\prime} \! \! \frac{d^3p}
 {(2\pi)^3}\frac{\tilde{p}_3^2D_0(\bm{p})}
 {E(\tilde{\bm{p}})[(g_3/v_{\perp})^2\tilde{\bm{p}}_{\perp}^2+(g_{\perp}/v_3)^2\tilde{p}_3^2]} \\
 \! \! &=& \! \! -\frac{\xi g_{\perp}^2}{v^2_{\perp}v_3^3} \! \int^{\prime} \! \! \frac{d^3\tilde{p}}
 {(2\pi)^3}\frac{\tilde{p}_3^2D_0(\bm{p})}
 {E(\tilde{\bm{p}})[(g_3/v_{\perp})^2\tilde{\bm{p}}_{\perp}^2+(g_{\perp}/v_3)^2\tilde{p}_3^2]} \\
 \! \! &=& \! \! \xi\beta\alpha lG_3(\beta,\delta)+O(l^2) \ , 
\end{eqnarray*}
\begin{eqnarray*}
 \Sigma^{(1)}_{\xi 4} \! \! &=& \! \! -\frac{2g_3^4}{v_{\perp}^4} \! \int^{\prime} \! \! \frac{d^3p}
 {(2\pi)^3}\frac{(\tilde{p}_2^2-\tilde{p}_1^2)\tilde{p}_2^2D_0(\bm{p})}
 {E(\tilde{\bm{p}})[(g_3/v_{\perp})^2\tilde{\bm{p}}_{\perp}^2+(g_{\perp}/v_3)^2\tilde{p}_3^2]^2} \\
 \! \! &=& \! \! \frac{2g_{\perp}^2g_3^6}{v_{\perp}^6v_3} \! \int^{\prime} \! \! \frac{d^3\tilde{p}}
 {(2\pi)^3}\frac{(\tilde{p}_2^2-\tilde{p}_1^2)\tilde{p}_2^2}
 {E(\tilde{\bm{p}})[(g_3/v_{\perp})^2\tilde{\bm{p}}_{\perp}^2+(g_{\perp}/v_3)^2\tilde{p}_3^2]^3} \\
 \! \! &=& \! \! \frac{g_{\perp}^2g_3^6}{8v_{\perp}^6v_3\pi^2} \! \int^{\Lambda}_{\Lambda/s} \! 
 d\tilde{p}_{\perp}\tilde{p}_{\perp} \! \int^{+\infty}_{-\infty} \! d\tilde{p}_3 \\
 \! \! & & \! \! \times\frac{\tilde{p}_{\perp}^4}
 {[(g_3/v_{\perp})^2\tilde{p}_{\perp}^2+(g_{\perp}/v_3)^2\tilde{p}_3^2]^3
 \sqrt{\tilde{p}_3^2+\tilde{p}_{\perp}^2+\tilde{b}^2\tilde{p}_{\perp}^4}} \\
 \! \! &=& \! \! \beta^3\alpha lG_4(\beta,\delta)+O(l^2) \ , 
\end{eqnarray*}
\begin{eqnarray*}
 \Sigma^{(2)}_{\xi 4} \! \! &=& \! \! -\frac{2\tilde{b}g_{\perp}^4}{v_3^4} \! \int^{\prime} \! \! 
 \frac{d^3p}{(2\pi)^3}\frac{(\tilde{p}_2^2-\tilde{p}_1^2)\tilde{p}_3^2D_0(\bm{p})}
 {E(\tilde{\bm{p}})[(g_3/v_{\perp})^2\tilde{\bm{p}}_{\perp}^2+(g_{\perp}/v_3)^2\tilde{p}_3^2]^2} \\
 \! \! & & \! \! +\frac{\tilde{b}g_{\perp}^4}{2v_3^4} \! \int^{\prime} \! \! \frac{d^3p}{(2\pi)^3}
 \frac{(\tilde{p}_2^2-\tilde{p}_1^2)D_0(\bm{p})}
 {E(\tilde{\bm{p}})[(g_3/v_{\perp})^2\tilde{\bm{p}}_{\perp}^2+(g_{\perp}/v_3)^2\tilde{p}_3^2]} \\
 \! \! &=& \! \! 0 \ ,
\end{eqnarray*}
and
\begin{eqnarray*}
 \Sigma^{(1)}_{\xi 5} \! \! &=& \! \! -\frac{2g_3^4}{v_{\perp}^4} \! \int^{\prime} \! \! \frac{d^3p}
 {(2\pi)^3}\frac{2\tilde{p}^2_1\tilde{p}_2^2D_0(\bm{p})}
 {E(\tilde{\bm{p}})[(g_3/v_{\perp})^2\tilde{\bm{p}}_{\perp}^2+(g_{\perp}/v_3)^2\tilde{p}_3^2]^2} \\
 \! \! &=& \! \! \frac{2g_{\perp}^2g_3^6}{v_{\perp}^6v_3} \! \int^{\prime} \! \! \frac{d^3\tilde{p}}
 {(2\pi)^3}\frac{2\tilde{p}^2_1\tilde{p}_2^2}
 {E(\tilde{\bm{p}})[(g_3/v_{\perp})^2\tilde{\bm{p}}_{\perp}^2+(g_{\perp}/v_3)^2\tilde{p}_3^2]^3} \\
 \! \! &=& \! \! \frac{g_{\perp}^2g_3^6}{8v_{\perp}^6v_3\pi^2} \! \int^{\Lambda}_{\Lambda/s} \! 
 d\tilde{p}_{\perp}\tilde{p}_{\perp} \! \int^{+\infty}_{-\infty} \! d\tilde{p}_3 \\
 \! \! & & \! \! \times\frac{\tilde{p}_{\perp}^4}
 {[(g_3/v_{\perp})^2\tilde{p}_{\perp}^2+(g_{\perp}/v_3)^2\tilde{p}_3^2]^3
 \sqrt{\tilde{p}_3^2+\tilde{p}_{\perp}^2+\tilde{b}^2\tilde{p}_{\perp}^4}} \\
 \! \! &=& \! \! \Sigma^{(1)}_{\xi 4} \ , 
\end{eqnarray*}
\begin{eqnarray*}
 \Sigma^{(2)}_{\xi 5} \! \! &=& \! \! -\frac{2\tilde{b}g_{\perp}^4}{v_3^4} \! \int^{\prime} \! \!
 \frac{d^3p}{(2\pi)^3}\frac{2\tilde{p}_1\tilde{p}_2\tilde{p}_3^2D_0(\bm{p})}
 {E(\tilde{\bm{p}})[(g_3/v_{\perp})^2\tilde{\bm{p}}_{\perp}^2+(g_{\perp}/v_3)^2\tilde{p}_3^2]^2} \\
 \! \! & & \! \! +\frac{\tilde{b}g_{\perp}^4}{2v_3^4} \! \int^{\prime} \! \! \frac{d^3p}{(2\pi)^3}
 \frac{2\tilde{p}_1\tilde{p}_2D_0(\bm{p})}
 {E(\tilde{\bm{p}})[(g_3/v_{\perp})^2\tilde{\bm{p}}_{\perp}^2+(g_{\perp}/v_3)^2\tilde{p}_3^2]} \\
 \! \! & & \! \! 0 \ ,
\end{eqnarray*}
where $\beta=\eta/r^2$ and
\begin{eqnarray*}
 G_{\perp}(\beta,z) \! \! &=& \! \! \! \int^{+\infty}_0 \! \frac{dt}{(t^2+\beta)^2\sqrt{t^2+1+z^2}} \ , 
 \\
 G_3(\beta,z) \! \! &=& \! \! \! \int^{+\infty}_0 \! dt\frac{2t^2}{(t^2+\beta)^2\sqrt{t^2+1+z^2}} \ , \\
 G_4(\beta,z) \! \! &=& \! \! \! \int^{+\infty}_0 \! \frac{dt}{(t^2+\beta)^3\sqrt{t^2+1+z^2}} \ .
\end{eqnarray*}

Finally, we calculate the one-loop contribution to the vertex function, which arises from $I_3$:
\begin{eqnarray*}
 I_3 \! \! &=& \! \! \frac{1}{6}\langle S_{int}^3\rangle_>+I_1I_2-\frac{1}{6}I_1^3=\frac{1}{6}\langle 
 S_{int}^3\rangle_> \\
 \! \! &=& \! \! i\Gamma_v \! \int_X\phi_<\rho_{0<}+\cdots \ ,
\end{eqnarray*}
where $\cdots$ denotes the terms with higher scaling dimensions. In view of $S_{int}$, $\Gamma_v$ is 
given by
\begin{eqnarray*}
 & & \! \! \Gamma_v \\
 & & \! \! =(-i)\frac{i^3}{6}\cdot 3\cdot 2 \! \sum_{\xi} \! \int^{\prime} \! \! \frac{d^3p}{(2\pi)^3} \! 
 \int^{+\infty}_{-\infty} \! \frac{dp_0}{2\pi}[-D_0(-\bm{p})] \\
 & & ~\times [G_{0\xi}(P)]^2 \\
 & & \! \! =- \! \sum_{\xi} \! \int^{\prime} \! \frac{d^3p}{(2\pi)^3}D_0(-\bm{p}) \! \int^{+\infty}_{-\infty} 
 \! \! \frac{dp_0}{2\pi}\frac{p_0^2-[E(\tilde{\bm{p}})]^2}{(p_0^2+[E(\tilde{\bm{p}})]^2)^2} \\
 & & \! \! =0 \ .
\end{eqnarray*}
As we have shown that $\Sigma_{\tau}=0$ to the one-loop order, this result indicates that our one-loop 
calculation is consistent with the Ward identity [Eq. (\ref{hodsmwt11})], and thus respects the gauge 
invariance.

Collecting the above results, $S_{eff}$ to the one-loop order is of the form
\begin{widetext}
\begin{eqnarray*}
 S_{eff}[\psi_<,\psi^{\dagger}_<,\phi_<] \! \! &=& \! \! \! \sum_{\xi} \! \int_X\psi^{\dagger}_{\xi<}
 \partial_{\tau}\psi_{\xi<}+[1+\beta^2\alpha lG_{\perp}(\beta,\delta)+O(l^2)]\sum_{\xi} \! \sum_{j=1,2} 
 \! \int_X\psi^{\dagger}_{\xi<}\Gamma_jv_{\perp}(-i\partial_j\psi_{\xi<}) \\
 \! \! & & \! \! +[1+\beta\alpha lG_3(\beta,\delta)+O(l^2)]\sum_{\xi} \! \int_X\psi^{\dagger}_{\xi<}
 \Gamma_3\xi v_3(-i\partial_3\psi_{\xi<}) \\
 \! \! & & \! \! +[1+\beta^3\alpha lG_4(\beta,\delta)+O(l^2)]\sum_{\xi} \! \sum_{j=4,5} \! \int_X
 \psi^{\dagger}_{\xi<}\Gamma_jbd_j(-i\bm{\nabla})\psi_{\xi<}+i \! \int_X\phi_<\rho_{0<} \\
 \! \! & & \! \! +\frac{1}{2} \! \int_X \! \left\{\! \left[\frac{1}{g_{\perp}^2}+\frac{l}{2v_3\pi^2}
 F_{\perp}(\delta)+O(l^2)\right] \! |\bm{\nabla}_{\perp}\phi_<|^2+ \! \left[\frac{1}{g_3^2}+\frac{l}
 {2rv_{\perp}\pi^2}F_3(\delta)+O(l^2)\right] \! |\partial_3\phi_<|^2\right\} \! +\cdots \ ,
\end{eqnarray*}
where $\cdots$ denotes the terms with higher scaling dimensions. Now we perform the scaling 
transformation
\begin{eqnarray*}
 x(y)\rightarrow e^lx(y) \ , ~z\rightarrow e^{z_3l}z \ , ~\tau\rightarrow e^{zl}\tau \ , ~
 \psi_{\xi<}=Z_{\psi}^{-1/2}\psi_{\xi} \ , ~\phi_<=Z_{\phi}^{-1/2}\phi \ ,
\end{eqnarray*}
to bring the terms $\int_X\psi^{\dagger}_{\xi<}\partial_{\tau}\psi_{\xi<}$ and $i\! \int_X\phi_<\rho_{0<}$ 
back to their original forms, yielding $Z_{\psi}=e^{(2+z_3)l}$ and $Z_{\phi}=e^{2zl}$. Therefore, 
$S_{eff}$ becomes
\begin{eqnarray*}
 S_{eff}[\psi,\psi^{\dagger},\phi] \! \! &=& \! \! \! \sum_{\xi} \! \int_X\psi^{\dagger}_{\xi}
 \partial_{\tau}\psi_{\xi}+e^{(z-1)l}[1+\beta^2\alpha lG_{\perp}(\beta,\delta)+O(l^2)]\sum_{\xi} \! 
 \sum_{j=1,2} \! \int_X\psi^{\dagger}_{\xi}\Gamma_jv_{\perp}(-i\partial_j\psi_{\xi}) \\
 \! \! & & \! \! +e^{(z-z_3)l}[1+\beta\alpha lG_3(\beta,\delta)+O(l^2)]\sum_{\xi} \! \int_X
 \psi^{\dagger}_{\xi}\Gamma_3\xi v_3(-i\partial_3\psi_{\xi}) \\
 \! \! & & \! \! +e^{(z-2)l}[1+\beta^3\alpha lG_4(\beta,\delta)+O(l^2)]\sum_{\xi} \! \sum_{j=4,5} \! 
 \int_X\psi^{\dagger}_{\xi}\Gamma_jbd_j(-i\bm{\nabla})\psi_{\xi} \\
 \! \! & & \! \! +\frac{1}{2}e^{(z_3-z)l} \! \left[\frac{1}{g_{\perp}^2}+\frac{l}{2v_3\pi^2}F_{\perp}
 (\delta)+O(l^2)\right] \! \int_X|\bm{\nabla}_{\perp}\phi|^2 \\
 \! \! & & \! \! +\frac{1}{2}e^{(2-z-z_3)l} \! \left[\frac{1}{g_3^2}+\frac{l}{2rv_{\perp}\pi^2}F_3
 (\delta)+O(l^2)\right] \! \! \int_X|\partial_3\phi|^2+i \! \int_X\phi\rho_0+\cdots \ .
\end{eqnarray*}
Consequently, the one-loop recursion relations for the parameters $v_{\perp}$, $v_3$, $b$, $g_{\perp}^2$, 
and $g_3^2$ are
\begin{eqnarray*}
 & & \! \! \frac{v_{\perp}^{\prime}}{v_{\perp}}=e^{(z-1)l}[1+\beta^2\alpha lG_{\perp}(\beta,\delta)
 +O(l^2)]=1+(z-1)l+\beta^2\alpha lG_{\perp}(\beta,\delta)+O(l^2) \ , \\
 & & \! \! \frac{v_3^{\prime}}{v_3}=e^{(z-z_3)l}[1+\beta\alpha lG_3(\beta,\delta)+O(l^2)]=1+(z-z_3)l
 +\beta\alpha lG_3(\beta,\delta)+O(l^2) \ , \\
 & & \! \! \frac{b^{\prime}}{b}=e^{(z-2)l}[1+\beta^3\alpha lG_4(\beta,\delta)+O(l^2)]=1+(z-2)l
 +\beta^3\alpha lG_4(\beta,\delta)+O(l^2) \ , \\
 & & \! \! \frac{g_{\perp}^2}{(g_{\perp}^{\prime})^2}=e^{(z_3-z)l}[1+2\alpha lF_{\perp}(\delta)+O(l^2)] 
 \! =1-(z-z_3)l+2\alpha lF_{\perp}(\delta)+O(l^2) \ , \\
 & & \! \! \frac{g_3^2}{(g_3^{\prime})^2}=e^{(2-z-z_3)l}[1+2\beta\alpha lF_3(\delta)+O(l^2)]
 =1-(z+z_3-2)l+2\beta\alpha lF_3(\delta)+O(l^2) \ .
\end{eqnarray*}
We choose the values of $z$ and $z_3$ such that $v_3$ and $b$ are both RG invariants, which leads to
Eq. (\ref{hodsmrg11}). Therefore, the one-loop recursion relations for the other parameters become
\begin{eqnarray*}
 \frac{r^{\prime}}{r} \! \! &=& \! \! 1+[1+\beta^2\alpha G_{\perp}(\beta,\delta)-\beta^3\alpha G_4
 (\beta,\delta)]l+O(l^2) \ , \\
 \frac{\delta^{\prime}}{\delta} \! \! &=& \! \! 1-[3+2\beta^2\alpha G_{\perp}(\beta,\delta)
 -2\beta^3\alpha G_4(\beta,\delta)]l+O(l^2) \ , \\
 \frac{g_{\perp}^2}{(g_{\perp}^{\prime})^2} \! \! &=& \! \! 1+[\beta\alpha G_3(\beta,\delta)+2\alpha 
 F_{\perp}(\delta)]l+O(l^2) \ , \\
 \frac{g_3^2}{(g_3^{\prime})^2} \! \! &=& \! \! 1-[2+\beta\alpha G_3(\beta,\delta)-2\beta^3\alpha G_4
 (\beta,\delta)-2\beta\alpha F_3(\delta)]l+O(l^2) \ .
\end{eqnarray*}
\end{widetext}
Note that 
$\delta^{\prime}=b^{\prime}\Lambda^{\prime}/(v_{\perp}^{\prime})^2=[b^{\prime}/(v_{\perp}^{\prime})^2]\Lambda e^{-l}$.
From these one-loop recursion relations, we obtain Eqs. (\ref{hodsmrg17}) -- (\ref{hodsmrg14}).



\begin{thebibliography}{99}
\bibitem{foot1} Without breaking the P and T symmetries, we can add a constant to $g_3$: 
        $g_3(\bm{k})=-t_3(C_3-\gamma+2-C_1-C_2)$, where $\gamma$ is real. For $|\gamma|<1$, there are 
        still two Dirac points located at $(0,0,\pm k_0)$ where $k_0=\cos^{-1}{\gamma}$.
\bibitem{foot2} The first BZ of this model consists of $2$D band structures parametrized by $k_z$. When 
        $b=0$, the T symmetry is preserved, and these $2$D band structures are divided into two phases:
        trivial insulators and the quantum spin Hall insulators (QSHIs), separated by the two planes with 
        $k_z=\pm k_0$. For a sample with open boundaries in the $x$ or $y$ directions, there are double 
        Fermi arcs terminated at the projections of the two Dirac nodes on the surface BZ. This is the 
        first-order DSM. When $b\neq 0$, the T symmetry is broken, and thus the edge modes are gapped. 
        The region occupied by trivial insulators remains intact, while the QSHIs become second-order TIs. 
        For a cube-shaped material, there will be gapless hinge modes at the hinges parallel to the 
        $z$-axis and terminated at the projections of the two Dirac nodes on the surface BZ. This is the 
        second-order DSM.
\end{thebibliography}
\end{document}